\documentclass[font=12]{article}

\usepackage{arxiv}
\usepackage[utf8]{inputenc} 
\usepackage[T1]{fontenc}    
\usepackage[hidelinks]{hyperref}       
\usepackage{url}            
\usepackage{booktabs}       
\usepackage{nicefrac}       
\usepackage{microtype}      
\usepackage{lipsum}		
\usepackage{graphicx}
\usepackage{doi}
\usepackage{comment}
\usepackage{caption}
\usepackage{subcaption}
\usepackage{bbm}
\usepackage{xcolor}
\usepackage{textcomp}

\usepackage{amsmath} 
\usepackage{amsfonts}       
\usepackage{amssymb}

\title{Locally Adaptive Conformal Prediction for Polynomial Chaos Expansions}

\date{} 					

\author{ \href{https://orcid.org/0000-0003-1264-1182}{\includegraphics[scale=0.06]{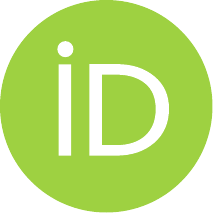}\hspace{1mm}Dimitrios Loukrezis}\\
	Scientific Computing Group\\
	Centrum Wiskunde \& Informatica\\
	Amsterdam, The Netherlands \\
	\texttt{d.loukrezis@cwi.nl} \\
    \And
	\href{https://orcid.org/0009-0004-3531-3119}
    {\includegraphics[scale=0.06]{orcid.pdf}\hspace{1mm} Arthur Hatstatt} \\
	Chair of Risk, Safety and Uncertainty Quantification\\
	ETH Z\"urich\\
	Z\"urich, Switzerland \\
	\texttt{arthur.hatstatt@ibk.baug.ethz.ch} \\
    \And
	\href{https://orcid.org/0000-0002-9501-7395}
    {\includegraphics[scale=0.06]{orcid.pdf}\hspace{1mm}Bruno Sudret} \\
	Chair of Risk, Safety and Uncertainty Quantification\\
	ETH Z\"urich\\
	Z\"urich, Switzerland \\
	\texttt{sudret@ethz.ch} \\
	\And
	\href{https://orcid.org/0000-0003-2272-2584}{\includegraphics[scale=0.06]{orcid.pdf}\hspace{1mm}Dimitris G.~Giovanis} \\
	Dpt. of Civil \& Systems Engineering\\
    Dpt. of Applied Mathematics and Statistics\\
	Johns Hopkins University\\
	Baltimore, MD, US \\
	\texttt{dgiovan1@jhu.edu} \\
}

\renewcommand{\headeright}{}
\renewcommand{\undertitle}{}

\usepackage[acronym]{glossaries}
\newacronym{cp}{CP}{conformal prediction}
\newacronym{cv}{CV}{cross validation}
\newacronym[longplural=Gaussian processes]{gp}{GP}{Gaussian process}
\newacronym{iid}{iid}{independent and identically distributed}
\newacronym{lar}{LAR}{least angle regression}
\newacronym{loo}{LOO}{leave-one-out}
\newacronym{ml}{ML}{machine learning}
\newacronym{ols}{OLS}{ordinary least squares}
\newacronym{omp}{OMP}{orthogonal matching pursuit}
\newacronym{pce}{PCE}{polynomial chaos expansion}
\newacronym{pdf}{PDF}{probability density function}
\newacronym{uq}{UQ}{uncertainty quantification}

\begin{document}
\maketitle

\begin{abstract}
Polynomial chaos expansions are widely used as surrogate models, but do not provide metrics that quantify the uncertainty in their predictions.
Conformal prediction offers such metrics in the form of prediction intervals with finite-sample coverage guarantees under the sole assumption of data exchangeability.
Existing combinations of conformal prediction with polynomial chaos expansions do not adapt the prediction interval width to the local prediction uncertainty, although the prediction error of a surrogate is typically input-dependent.
To address this limitation, this work develops a locally adaptive jackknife+ conformal prediction framework for polynomial chaos expansions. 
Therein, the non-conformity scores are normalized with an auxiliary polynomial chaos expansion trained on the log-transformed leave-one-out residuals of the base model.
We show that the coverage guarantee of jackknife+ conformal prediction carries over to the adaptive intervals if the leave-one-out auxiliary models are trained on nested leave-one-out residuals.
For fixed-basis polynomial chaos expansions fitted by ordinary least squares, all required quantities are derived in closed form and the guarantee-preserving adaptive intervals are obtained from a single model fit, without any model retraining.
For sparse polynomial chaos expansions, where multiple model fits are necessary, we propose a heuristic simplification that limits the computational overhead of adaptivity at the cost of the formal coverage guarantee.
Numerical experiments on an academic benchmark with known heteroscedastic noise and on an engineering simulation dataset show that the suggested method attains the target coverage while providing intervals that adapt to local prediction uncertainty, therefore being considerably more statistically efficient. 
These advantages are particularly prominent in high-uncertainty regions of the input space, where the performance of non-adaptive methods degrades substantially.
\end{abstract}


\section{Introduction}
\label{sec:intro}
Surrogate models have become indispensable for enabling and accelerating parametric studies that require repeated model evaluations, for example, by replacing computationally expensive simulation models in time- and resource-demanding studies. 
Example surrogate-assisted tasks include optimization~\cite{jiang2019surrogate}, \gls{uq}~\cite{owen2017comparison}, or combined tasks such as optimization under uncertainty~\cite{moustapha2019surrogate, moustapha2022multi}. 
Owing to the advances in the field of \gls{ml}, data-driven surrogates based on \gls{ml} techniques have enjoyed increased popularity in recent years~\cite{koziel2022fundamentals, chen2025machine}. 
However, the predictive capability of a data-driven surrogate model is governed by a number of factors, such as its model form and the available training data. 
Its predictions are therefore subject to uncertainty, for example due to approximation errors or the intrinsic variability of the observed output in the case of noisy observations.
For surrogate-based results to be trustworthy, prediction uncertainty must be quantified reliably.

Among the numerous available surrogate modeling techniques, \glspl{pce}~\cite{xiu2002wiener, ghanem2003stochastic} are particularly popular in the field of \gls{uq}, because their spectral structure gives direct access to statistical moments and sensitivity indices~\cite{sudret2008global}.
Moreover, regression-based \glspl{pce} are commonly employed as data-driven \gls{ml} models~\cite{torre2019data}. 
The accuracy of a \gls{pce} is typically assessed with global generalization error metrics, for example based on the \gls{loo} error~\cite{blatman2011adaptive} or using a separate validation dataset.
However, these metrics provide a single number for the entire input space and do not indicate how reliable an individual prediction is.
In contrast to probabilistic surrogates such as \glspl{gp}~\cite{rasmussen2005gaussian} or Bayesian neural networks~\cite{arbel2026primer}, standard \glspl{pce} do not provide metrics that quantify the model's predictive uncertainty, such as prediction, confidence, or credibility intervals.
Such intervals can be obtained by combining \glspl{pce} with bootstrapping~\cite{marelli2018active} or ensemble modeling~\cite{loukrezis2022power, shang2024active}. 
However, the spread of bootstrap or ensemble predictions mainly reflects the variability of the estimated model with respect to the training data, largely neglecting the impact of noise or systematic approximation errors.
Other options include hybrid \gls{gp}-\gls{pce} models~\cite{schobi2015polynomial, kersaudy2015new, manfredi2025hybrid, manfredi2026polynomial} or Bayesian \gls{pce} formulations~\cite{shao2017bayesian, li2025new, rumsey2026bayesian}, where interval validity largely depends on the underlying modeling assumptions, e.g., covariance kernel, prior distributions, or noise model.
Moreover, none of these approaches guarantees that the resulting intervals contain the true output with a prescribed probability for a finite amount of data.

\Gls{cp} methods~\cite{vovk2005algorithmic} offer such guarantees and can additionally be combined with any regression model.
Under the assumption of data exchangeability and for any finite number of data points, \gls{cp} produces prediction intervals with marginal coverage guarantees, meaning that, on average, the intervals contain the true output with a prescribed probability. 
This probability is commonly called the target coverage and denoted by $(1-\alpha)$, where $\alpha \in (0,1)$ is the corresponding significance level. 
The original full (or transductive) \gls{cp} variant~\cite{vovk2005algorithmic} uses all available data for both training and interval calibration, but requires retraining the regression model for each test input and for multiple candidate output values. 
This is computationally prohibitive in general and becomes tractable only for specific models~\cite{nouretdinov2001ridge}.
The most widely used variant is split (or inductive) \gls{cp}~\cite{papadopoulos2002inductive, lei2018distribution}, where the model is trained once and the intervals are calibrated on a held-out dataset.
Although computationally efficient, split \gls{cp} reduces the amount of data available for training, which can be detrimental to model accuracy if only limited data are available.
Cross \gls{cp}~\cite{vovk2015cross} mitigates this drawback by calibrating the prediction intervals via $k$-fold \gls{cv}, at the cost of training $k$ models and only approximate interval validity, i.e., without offering a coverage guarantee.
For the limit-case of $k=M$, $M$ being the size of the training dataset, cross \gls{cp} becomes jackknife \gls{cp}~\cite{lei2018distribution}, where \gls{loo} \gls{cv} is used for interval calibration.
In particular, the jackknife+ variant~\cite{barber2021predictive} offers a $(1-2\alpha)$ finite-sample coverage guarantee. 
While this is weaker than the $(1-\alpha)$ coverage guarantee provided by full and split \gls{cp}, it is still favorable compared to the only approximate interval validity of standard jackknife \gls{cp}.
Moreover, the empirical coverage of jackknife+ \gls{cp} is typically on par with the target coverage~\cite{barber2021predictive}.

To the best of our knowledge, two works have considered the combination of \gls{cp} with \glspl{pce} so far.
Loukrezis and Giovanis~\cite{loukrezis2026conformalized} combined jackknife and jackknife+ \gls{cp} with total-degree \glspl{pce} fitted by \gls{ols}. 
In this case, \gls{loo} residuals and \gls{loo} model predictions can be computed with closed-form expressions and the \gls{cp} intervals are obtained at practically no additional cost compared to the \gls{pce} fit itself.
Hatstatt et al.~\cite{hatstatt2026conformal} combined full and jackknife+ \gls{cp} with both total-degree \gls{ols}-fitted \glspl{pce} and sparse \glspl{pce} based on a hybrid \gls{lar} method~\cite{blatman2011adaptive}. 
In the former case, jackknife+ \gls{cp} intervals are again computed via closed-form expressions. 
Full \gls{cp} intervals are also computed with very low computational overhead, through an exact computation of the interval limits that exploits the linearity of the \gls{ols} solution with respect to the model outputs.
For sparse \glspl{pce}, Hatstatt et al.\ show that reusing the basis selected on the full training dataset violates the symmetry requirement of \gls{cp} and leads to a severe performance degradation.
The adaptation of full \gls{cp} to sparse \glspl{pce} was made possible by exploiting the linearity of the \gls{lar} solution path and solving equations to determine the prediction interval bounds. 
In contrast, jackknife+ \gls{cp} was adapted through repeated \gls{loo} refitting of the sparse \gls{pce} model, taking advantage of the considered small-data regime, for which the computational cost remains manageable.

The aforementioned \gls{cp}-\gls{pce} combinations use absolute residuals as non-conformity scores, which measure the discrepancy between the model prediction and the true output.
For jackknife or jackknife+ \gls{cp}, this results in intervals of nearly constant width over the entire input space.
In contrast, the prediction error of a surrogate model is often heteroscedastic, i.e., input-dependent. 
For example, this is the case in settings with input-dependent noise, localized nonlinearities, or unevenly distributed training data.
Full \gls{cp} intervals do vary in width, however, this variation stems only from the position of the test input relative to the training data, similar to a \gls{gp} with stationary kernel and homoscedastic noise.
In either case, intervals that do not account for the local prediction uncertainty of the model are typically too conservative in regions of the input space characterized by low predictive uncertainty and, conversely, too narrow in high-uncertainty regions.
The target coverage may still be satisfied on average by over-covering low-uncertainty outputs, while simultaneously under-covering high-uncertainty ones, which however is not the desired behavior. 
In fact, reliable prediction intervals are needed primarily for regions of the input space where the prediction uncertainty is high.

Locally adaptive \gls{cp} addresses this issue by normalizing the non-conformity scores with an estimate of the local prediction uncertainty~\cite{papadopoulos2008normalized, papadopoulos2011reliable, deutschmann2024adaptive}, commonly obtained with an auxiliary regression model trained on the residuals of the base model.
Alternatively, adaptive intervals can be obtained by conformalized quantile regression models~\cite{romano2019conformalized}.
For split \gls{cp}, non-conformity score normalization is straightforward, as long as the auxiliary model is trained independently of the calibration data.
For jackknife-based \gls{cp}, the auxiliary model must be trained in a \gls{loo} fashion as well. 
As also discussed in this work, simply training it on the \gls{loo} residuals of the base model violates the symmetry requirement on which the coverage guarantee of jackknife+ \gls{cp} relies.
Interval adaptivity is considerably more demanding for the full \gls{cp} variant~\cite{gazin2024transductive}, where the exact computation of the interval limits exploited by Hatstatt et al.~\cite{hatstatt2026conformal} no longer applies.
Due to its impractical computational cost, adaptive full \gls{cp} is not pursued in this work.
An adaptive variant of the standard jackknife \gls{cp} method is also not pursued, as it offers no finite-sample coverage guarantee and is typically less robust than the jackknife+ variant~\cite{barber2021predictive}.

To address the aforementioned limitations, this work develops an adaptive jackknife+ \gls{cp} framework for \gls{pce} regression, which complements \gls{pce} predictions with locally adaptive prediction intervals.
Within the suggested approach, the local uncertainty estimator utilizes an auxiliary \gls{pce} model trained on the log-transformed \gls{loo} residuals of the base \gls{pce}.
We show that the $(1-2\alpha)$ coverage guarantee of jackknife+ \gls{cp} transfers to the adaptive intervals if the \gls{loo} auxiliary models are trained on nested \gls{loo} residuals, i.e., residuals of base models that omit two training data points at a time.
For fixed-basis, \gls{ols}-fitted \glspl{pce}, the nested \gls{loo} residuals and the \gls{loo} auxiliary model predictions can be computed with closed-form expressions. 
The guarantee-preserving adaptive \gls{cp} intervals are then obtained from a single \gls{ols} fit of the base model, without any model retraining.
For sparse \glspl{pce}, however, the guarantee-preserving approach becomes computationally prohibitive. 
To mitigate the computational demand, we suggest a heuristic simplification that limits the overhead substantially, but forgoes the formal coverage guarantee. 
Nonetheless, we demonstrate empirically that the resulting models attain the target coverage, at least in the use cases considered in this work.

The suggested adaptive jackknife+ \gls{cp}-\gls{pce} method is compared against its non-adaptive counterpart, as well as against adaptive and non-adaptive split \gls{cp}-\glspl{pce}, first on a benchmark with known heteroscedastic noise, and subsequently on an engineering simulation dataset.
The numerical results show clearly that adaptive jackknife+ \gls{cp}-\glspl{pce} attain the prescribed coverage levels, while simultaneously adapting the prediction intervals according to the local prediction uncertainty. 
Importantly, adaptive jackknife+ \gls{cp}-\glspl{pce} attain an empirical coverage close to the target level also in the high-uncertainty regions of the input space, where the coverage of the non-adaptive variants drops substantially.
Moreover, adaptive jackknife+ \gls{cp}-\glspl{pce} perform on par with adaptive split \gls{cp}-\glspl{pce} calibrated on very large datasets, and outperform them when the available data must be shared between training and calibration.

The remainder of this paper is organized as follows.
Section~\ref{sec:cp} recalls the main components of \gls{cp} for regression models, focusing on the \gls{cp} variants employed in this work.
Section~\ref{sec:pce} discusses briefly the regression-based \gls{pce} method.
Section~\ref{sec:acp-pce} presents the suggested methodology for integrating jackknife+ \gls{cp} into \glspl{pce}, so that their predictions are complemented with locally adaptive, input-dependent \gls{cp} intervals.
Numerical experiments demonstrating the applicability of the method to both an academic benchmark and an engineering use case are presented in Section~\ref{sec:num-exp}.
Finally, the findings of this work are summarized in Section~\ref{sec:conclusion}.

\section{Conformal prediction}
\label{sec:cp}

\subsection{Conformal prediction preliminaries}
\label{sec:cp-prelim}
We consider a regression model $\widehat{\mu}$ trained on a dataset $\mathcal{D}_{\text{train}} = \left\{\left(\mathbf{x}^{(m)}, y^{(m)}\right)\right\}_{m=1}^M$, where $\mathbf{x}^{(m)} \in \mathbb{R}^N$ is an input vector and $y^{(m)} \in \mathbb{R}$ is the corresponding scalar output\footnote{We only consider scalar outputs in this work. For vector outputs, the element-wise application of \gls{cp} provides element-wise coverage only, and dedicated techniques are needed for joint \gls{cp} regions~\cite{dheur2025unified}. These techniques are out of scope for the present work.}. 
Given a test point $\left( \mathbf{x}^*, y^*\right) \not\in \mathcal{D}_{\text{train}}$, the regression model produces a prediction $\widehat{y}^* = \widehat{\mu}(\mathbf{x}^*)$ for the true output $y^*$.
\gls{cp} aims to construct a prediction interval $\widehat{C}_{1-\alpha}\left(\mathbf{x}^*\right)$, where $\alpha \in \left(0,1\right)$ denotes a pre-defined significance level, that satisfies the condition
\begin{equation}
\label{eq:coverage-condition}
\mathbb{P}\left\{y^* \in \widehat{C}_{1-\alpha}\left(\mathbf{x}^*\right)\right\} \geq 1 - \alpha,
\end{equation}
meaning that the interval contains the true output with a target coverage probability of $(1-\alpha)$.
The probability is taken with respect to the joint distribution of the data used to calibrate the interval (see Section~\ref{sec:non-conformity-scores}) and the test data. 
Therefore, the provided coverage is marginal, not conditional.
In fact, exact conditional coverage is in general impossible to achieve without strong distributional assumptions~\cite{vovk2012conditional, barber2021limits}. 

A key assumption of \gls{cp} is data exchangeability. 
Specifically, the joint distribution of the data used for interval calibration and the test data must be invariant under arbitrary sample permutations. 
In practice, exchangeability is typically satisfied when the data points are \gls{iid}, although it is formally a weaker assumption.
For \gls{cp} variants where the training data enter the interval calibration process, for example, the jackknife \gls{cp} variant presented in Section~\ref{sec:jackknife-cp}, an additional requirement is permutation symmetry of the learning algorithm, meaning that permutations in the training data must not alter the resulting regression model and its predictions.

\subsection{Non-conformity scores}
\label{sec:non-conformity-scores}
\Gls{cp} quantifies the discrepancy between a data point and the regression model's prediction through a non-conformity measure $s\left(\mathbf{x},y;\widehat{\mu}\right)$~\cite{kato2023review}, which is evaluated on a set of data referred to as the calibration data.
Each evaluation produces a scalar value called the non-conformity score, where
larger scores indicate samples that are less consistent with the model predictions.

Let $\widehat{q}_{1-\alpha}$ denote the $(1-\alpha)$-quantile of the empirical distribution of a set of non-conformity scores $\left(s_j\right)_{j=1}^J$.
This quantile is subsequently used to define the prediction interval for a test input $\mathbf{x}^*$. 
The computation of the quantile, the construction of the prediction interval, as well as which data provide the non-conformity scores, depend on the specific \gls{cp} variant (see Section~\ref{sec:cp-variants}).

For regression tasks, it is common to use the absolute residuals as non-conformity scores:
\begin{equation}
\label{eq:abs-residual}
s_j = R_j = \left|r_j\right| = \left| y^{(j)} - \widehat{\mu}\left(\mathbf{x}^{(j)}\right) \right|,
\end{equation}
where $r_j$ denotes the signed residual.
For the split and jackknife+ \gls{cp} variants considered in this work, this leads to prediction intervals whose widths remain constant or nearly constant across the input space and do not account for local variations due to input‑dependent model accuracy, e.g., because of heteroscedastic noise or locally varying model error.
This limitation motivates normalized non-conformity scores~\cite{papadopoulos2008normalized, deutschmann2024adaptive}, given as 
\begin{equation}
\label{eq:normalized-score}
s^{\text{n}}_j =
\frac{R_j}{\widehat{\eta}\left(\mathbf{x}^{(j)}\right)},
\end{equation}
where $\widehat{\eta}\left( \mathbf{x} \right) > 0$ is a function that estimates the local prediction uncertainty for the given input.
This residual scaling results in prediction intervals whose width follows
$\widehat{\eta}$, thus also in \emph{adaptive} intervals, i.e., intervals whose width varies across the input space with the local magnitude of the prediction error, to the extent that $\widehat{\eta}$ tracks this magnitude.
Note that a varying interval width alone does not imply adaptivity.
For example, the widths of full \gls{cp} intervals with non-conformity scores as in Eq.~\eqref{eq:abs-residual} vary with the location of the test input relative to the training inputs, thus not reflecting input-dependent noise or model error.

\subsection{Conformal prediction variants}
\label{sec:cp-variants}

In the following, we outline the two main variants of the \gls{cp} method that are of interest in this work, namely, split \gls{cp} and jackknife \gls{cp}.
The former is used as the baseline validation method, whereas the latter is combined with \gls{pce} in the adaptive \gls{cp}-\gls{pce} framework presented in Section~\ref{sec:acp-pce}.

\subsubsection{Split conformal prediction}
\label{sec:split-cp}
Split or inductive \gls{cp}~\cite{papadopoulos2002inductive} is the most commonly used \gls{cp} method due to its simplicity and efficiency in terms of model training and interval calibration. 
The regression model $\widehat{\mu}$ is fitted once on $\mathcal{D}_{\text{train}}$ and then held fixed.
A disjoint calibration dataset $\mathcal{D}_{\text{cal}} = \left\{\left(\mathbf{x}^{(j)}, y^{(j)}\right)\right\}_{j=1}^J$
is used to compute the non‑conformity scores $s_j = s\left(\mathbf{x}^{(j)}, y^{(j)};\widehat{\mu}\right)$.
The quantile $\widehat{q}_{1-\alpha}$ is set to the $k$-th order statistic of $(s_j)_{j=1}^J$ with $k = \lceil (J+1)(1-\alpha) \rceil$.
The prediction interval is then constructed as 
\begin{equation}
\label{eq:split-cp-interval-general}
\widehat C_{1-\alpha}\left(\mathbf{x}^*\right) \;=\; \left\{\, y : \; s\left(\mathbf{x}^*,y;\widehat{\mu}\right) \le \widehat q_{1-\alpha}\,\right\},
\end{equation}
where $\widehat{q}_{1-\alpha}=+\infty$ if $k>J$.
Using absolute residuals as non-conformity scores, the interval in Eq.~\eqref{eq:split-cp-interval-general} simplifies to 
\begin{equation}
\label{eq:split-cp-interval-abs-residual}
\widehat{C}_{1-\alpha}(\mathbf{x}^*) =
\left[\,\widehat{\mu}\left(\mathbf{x}^*\right) - \widehat{q}_{1-\alpha}, \; 
\widehat{\mu}\left(\mathbf{x}^*\right) + \widehat{q}_{1-\alpha}\,\right],
\end{equation}
which is symmetric and has constant width equal to $2\widehat{q}_{1-\alpha}$. 
Using normalized non-conformity scores results in input-dependent prediction intervals given as 
\begin{equation}
\label{eq:split-cp-interval-adaptive}
\widehat{C}_{1-\alpha}(\mathbf{x}^*) =
\left[\,\widehat{\mu}\left(\mathbf{x}^*\right) - \widehat{q}_{1-\alpha} \widehat{\eta}\left(\mathbf{x}^*\right), \; 
\widehat{\mu}\left(\mathbf{x}^*\right) + \widehat{q}_{1-\alpha} \widehat{\eta}\left(\mathbf{x}^*\right)\,\right],
\end{equation}
where the quantile $\widehat{q}_{1-\alpha}$ is now computed from the normalized non-conformity scores $\left(s_j^{\mathrm{n}}\right)_{j=1}^J$ defined in Eq.~\eqref{eq:normalized-score}.

Conditionally on the training set, the calibration and test points are assumed exchangeable.
Under this assumption, split \gls{cp} guarantees the finite‑sample marginal coverage condition in Eq.~\eqref{eq:coverage-condition} by construction, while requiring only a single model training and a single pass over the calibration data.
Its main disadvantage is the need for a disjoint calibration dataset, which increases data demand and can be problematic in small-data regimes.

\subsubsection{Jackknife-based conformal prediction}
\label{sec:jackknife-cp}
Jackknife-based \gls{cp} methods extend the idea of split \gls{cp} to \gls{loo} \gls{cv}. 
In the following, we differentiate between standard jackknife \gls{cp}~~\cite{lei2018distribution} and the jackknife+~\cite{barber2021predictive} variant of the method. 
In either case, the \gls{loo} models $\widehat{\mu}_{-m}$, $m=1, \dots,M$, are trained on the corresponding \gls{loo} datasets $\mathcal{D}_{-m} = \mathcal{D}_{\text{train}} \setminus \left\{\left(\mathbf{x}^{(m)}, y^{(m)}\right)\right\}$, and used to compute the absolute \gls{loo} residuals
\begin{equation}
\label{eq:loo-residuals}
R_m^{\text{LOO}} = \left| r_m^{\text{LOO}} \right| = \left| y^{(m)} - \widehat{\mu}_{-m}\left(\mathbf{x}^{(m)}\right) \right|,
\end{equation} 
which are used as non-conformity scores, i.e., $s_m = R_m^{\text{LOO}}$.
Normalized non-conformity scores can also be used, in this case given as
\begin{equation}
\label{eq:loo-residuals-normalized}
s_m^{\text{n}} = \frac{R_m^{\text{LOO}}}{\widehat{\eta}_{-m}\left(\mathbf{x}^{(m)}\right)},
\end{equation}
where $\widehat{\eta}_{-m}$ denotes a local uncertainty estimator fitted on $\mathcal{D}_{-m}$ only, similar to $\widehat{\mu}_{-m}$. 
While any type of local uncertainty estimator can be used in principle, the restriction to $\mathcal{D}_{-m}$ ensures that neither $\widehat{\mu}_{-m}$ nor $\widehat{\eta}_{-m}$ depends on the $m$-th training point, which is therefore treated in the same way as an out-of-sample data point. 
This is a formal requirement for the coverage guarantee of jackknife+ \gls{cp} (discussed next), but also a natural choice for standard jackknife \gls{cp} as well.
As discussed in Section~\ref{sec:acp-pce-eta}, this requirement is more subtle
than it appears, particularly when the uncertainty estimator is trained on \gls{loo} quantities.

For standard jackknife \gls{cp}, a final model $\widehat{\mu}$ and, in the normalized case, a final uncertainty estimator $\widehat{\eta}$ are trained on the full training dataset $\mathcal{D}_{\text{train}}$.
The prediction interval is then constructed similar to Eq.~\eqref{eq:split-cp-interval-abs-residual} when using the absolute \gls{loo} residuals, or similar to Eq.~\eqref{eq:split-cp-interval-adaptive} when using the normalized \gls{loo} residuals, with the quantile $\widehat{q}_{1-\alpha}$ computed from $\left(s_m\right)_{m=1}^M$ or $\left(s_m^{\text{n}}\right)_{m=1}^M$, respectively.
Jackknife \gls{cp} does not offer a finite‑sample coverage guarantee, however, its empirical coverage typically matches the target level.

Jackknife+ is an extension of jackknife \gls{cp} that achieves a $(1-2\alpha)$ finite‑sample marginal coverage guarantee by constructing the interval from quantiles of the \gls{loo} model predictions $\widehat{\mu}_{-m}\left(\mathbf{x}^*\right)$ shifted by the \gls{loo} residual-based non-conformity scores. 
The \gls{cp} interval of the non-adaptive jackknife+ variant is given as
\begin{equation}
\label{eq:jackknife-plus-interval-nonadaptive}
\widehat{C}_{1-\alpha}\left(\mathbf{x}^*\right) =
\left[\,
\widehat{q}^{\,-}_{\alpha}\left(\left\{ \widehat{\mu}_{-m}\left(\mathbf{x}^*\right)
  - s_m \right\}_{m=1}^{M}\right),\;
\widehat{q}^{\,+}_{1-\alpha}\left(\left\{ \widehat{\mu}_{-m}\left(\mathbf{x}^*\right)
  + s_m \right\}_{m=1}^{M}\right)
\,\right],
\end{equation}
where $\widehat{q}^{\,-}_{\alpha}$ and $\widehat{q}^{\,+}_{1-\alpha}$ denote the $\lfloor \alpha (M+1) \rfloor$-th and $\lceil (1-\alpha)(M+1) \rceil$-th smallest values of the respective sets, respectively corresponding to the $\alpha$- and $(1-\alpha)$-quantiles of these sets.
In the adaptive variant, the \gls{cp} interval is modified as
\begin{equation}
\label{eq:jackknife-plus-interval-adaptive}
\widehat{C}_{1-\alpha}\left(\mathbf{x}^*\right) =
\left[\,
\widehat{q}^{\,-}_{\alpha}\left(\left\{ \widehat{\mu}_{-m}\left(\mathbf{x}^*\right)
  - s_m^{\text{n}} \, \widehat{\eta}_{-m}\left(\mathbf{x}^*\right) \right\}_{m=1}^{M}\right),\;
\widehat{q}^{\,+}_{1-\alpha}\left(\left\{ \widehat{\mu}_{-m}\left(\mathbf{x}^*\right)
  + s_m^{\text{n}} \, \widehat{\eta}_{-m}\left(\mathbf{x}^*\right) \right\}_{m=1}^{M}\right)
\,\right],
\end{equation}
such that each normalized \gls{loo} residual is rescaled by the local uncertainty estimate $\widehat{\eta}_{-m}(\mathbf{x}^*)$ at the test input. 
Note that, in this case, it is not necessary to train a model $\widehat{\mu}$ and an uncertainty estimator $\widehat{\eta}$ on the full training dataset $\mathcal{D}_{\text{train}}$.

Jackknife-based \gls{cp} methods are more data-efficient compared to split \gls{cp}, as a single dataset is used for both model training and interval calibration. 
This comes at the cost of a relaxed coverage guarantee for jackknife+ \gls{cp}, or no guarantee for jackknife \gls{cp}. 
Moreover, $M+1$ (for jackknife) or $M$ (for jackknife+) regression models must be trained, compared to the single trained model in split \gls{cp}, which also extends to the local uncertainty estimator in case normalized non-conformity scores are employed.
However, this repeated training can be avoided for certain models, for which the \gls{loo} residuals and model predictions can be computed with analytical, closed-form expressions~\cite{loukrezis2026conformalized, hatstatt2026conformal, wang2018fast}, as also shown in Section~\ref{sec:acp-pce-fixed-ols}. 

Note that, despite their differences, the jackknife and jackknife+ \gls{cp} methods produce very similar prediction intervals in practice~\cite{barber2021predictive, loukrezis2026conformalized}. 
In the following, only the jackknife+ \gls{cp} method is employed, mainly because it provides a formal coverage guarantee and is typically the more robust variant between the two~\cite{barber2021predictive}.

\section{Polynomial chaos regression}
\label{sec:pce}
Similar to Section~\ref{sec:cp}, we consider input-output pairs $\left(\mathbf{x}, y\right)$, the relation of which is now given through a black-box model $\mu: \mathbf{x} \rightarrow y$, such that $y=\mu(\mathbf{x})$.
The inputs are assumed to be realizations $\mathbf{x} = \mathbf{X}(\omega) \in \Xi \subset \mathbb{R}^N$ of a random vector $\mathbf{X} = \left(X_1, \dots, X_N\right)$ with \gls{pdf} $f_{\mathbf{X}}: \Xi \rightarrow \mathbb{R}_{\geq 0}$, where $\omega$ denotes a possible outcome and $\Xi$ the input space.
The random variables $X_n$, $n=1,\dots,N$, are assumed to be mutually independent.
By propagation of uncertainty~\cite{rao2006probability}, $Y = \mu\left(\mathbf{X}\right)$ is a random variable dependent on $\mathbf{X}$. 

If $Y = \mu\left(\mathbf{X}\right)$ is square-integrable, it admits a representation in the form of a \gls{pce}~\cite{xiu2002wiener, ghanem2003stochastic}, such that 
\begin{equation}
 \label{eq:pce-infinite}
y = \mu \left(\mathbf{x} \right) =  \sum_{\mathbf{k} \in \mathbb{N}_0^N} c_{\mathbf{k}} \Psi_{\mathbf{k}} \left( \mathbf{x} \right), 
\end{equation}
where  $\Psi_{\mathbf{k}}$ are multivariate polynomials, $c_{\mathbf{k}} \in \mathbb{R}$ are the associated coefficients, and ${\mathbf{k}} = \left(k_1, \dots, k_N\right)$ are unique multi-indices defined by the polynomial's partial degree in each input dimension.
Following the Wiener-Askey scheme~\cite{xiu2002wiener} or by numerical construction~\cite{soize2004physical, oladyshkin2012data}, the polynomials form an orthogonal basis with respect to the input \gls{pdf}, thus satisfying 
\begin{equation}
\label{eq:orthogonality}
\mathbb{E}\left[\Psi_{\mathbf{k}} \Psi_{\mathbf{l}}\right] = \int_{\Xi}  \Psi_{\mathbf{k}}\left(\mathbf{x}\right) \Psi_{\mathbf{l}}\left(\mathbf{x}\right) f_{\mathbf{X}}\left(\mathbf{x}\right) \mathrm{d}\mathbf{x} =  \mathbb{E}\left[\Psi_{\mathbf{k}}^2\right] \delta_{\mathbf{k} \mathbf{l}},
\end{equation}
where $\delta_{\mathbf{k} \mathbf{l}}$ denotes the Kronecker delta.
An orthonormal basis is obtained for $\mathbb{E}\left[\Psi_{\mathbf{k}}^2\right] = 1$. 
By truncating the infinite series in Eq.~\eqref{eq:pce-infinite} to finite terms, a \gls{pce} approximation (or surrogate model) is obtained:
\begin{equation}
\label{eq:pce-truncated}
y \approx \widehat{\mu} \left(\mathbf{x} \right) = \sum_{k=1}^K \widehat{c}_{k} \Psi_{k} \left( \mathbf{x} \right) = \sum_{\mathbf{k} \in \Lambda} \widehat{c}_{\mathbf{k}} \Psi_{\mathbf{k}} \left( \mathbf{x} \right), 
\end{equation}
where $\Lambda \subset \mathbb{N}_0^N$ is a multi-index set with cardinality $\#\Lambda = K$ and there exists a one-to-one relation between $k=1,\dots,K$ and $\mathbf{k} \in \Lambda$. 
Basis truncation is commonly based on $q$-quasi-norms~\cite{blatman2011adaptive}, such that $\Lambda = \left\{ \mathbf{k}: \lVert \mathbf{k} \rVert_q = \left( \sum_{n=1}^N k_n^q \right)^{1/q} \leq P \right\}$, where $0< q \leq1$ and $P$ denotes the maximum allowed univariate polynomial degree.
For $q=1$, the so-called total-degree \gls{pce} with basis size $K = \frac{\left(N+P\right)!}{N! P!}$ is obtained.
Lower $q$-values exclude indices with high-order interactions, resulting in smaller bases, while still including univariate polynomials up to degree $P$.

In the regression setting,  a training dataset $\mathcal{D}_{\text{train}} = \left\{\left(\mathbf{x}^{(m)}, y^{(m)}\right)\right\}_{m=1}^M$ as in Section~\ref{sec:cp} is given, and the so-called design matrix $\mathbf{D} \in \mathbb{R}^{M \times K}$ is constructed such that $d_{mk} = \Psi_k\left( \mathbf{x}^{(m)} \right)$. 
Moreover, the model outputs are collected into a vector $\mathbf{y} = \left(y^{(1)}, \dots, y^{(M)}\right)^\top$.
The coefficients $\widehat{\mathbf{c}} = \left(\widehat{c}_1, \dots, \widehat{c}_K\right)^\top$ can then be estimated by solving the \gls{ols} problem
\begin{equation} 
\label{eq:regression-matrix}
\widehat{\mathbf{c}} = \underset {\mathbf{c} \in \mathbb{R}^K}{\arg\min} \left\{ \lVert \mathbf{y} - \mathbf{D} \widehat{\mathbf{c}} \rVert_2^2 \right\},
\end{equation}
and the \gls{pce} is in fact a linear regression model. 
Provided that $M \ge K$ and that $\mathbf{D}$ has full column rank, the \gls{ols} problem in Eq.~\eqref{eq:regression-matrix} admits the closed-form solution
\begin{equation} 
\label{eq:regression-solution}
\widehat{\mathbf{c}} = \left( \mathbf{D}^\top \mathbf{D}\right)^{-1} \mathbf{D}^\top \mathbf{y}.
\end{equation} 
 To ensure the stability and accuracy of the \gls{pce}, it is common to use training data sizes $M = C K$ with $C \in \left[2, 10\right]$~\cite{migliorati2014analysis}.
 
The number of terms in fixed-basis \glspl{pce}, where the basis is selected prior to fitting, grows rapidly for increasing input dimension $N$ and maximum polynomial degree $P$.
Combined with the data demand for reliably estimating the coefficients, this can lead to a severe computational bottleneck.
Motivated by the so-called sparsity-of-effects principle~\cite{blatman2009phd}, a common mitigation approach is to augment the regression problem \eqref{eq:regression-matrix} with an $\ell_1$-penalty term, leading to the sparsity-promoting optimization problem
\begin{equation}
\label{eq:regression-matrix-penalty}
\widehat{\mathbf{c}} = \underset{\mathbf{c} \in \mathbb{R}^K}{\arg\min}
\left\{
\left\lVert \mathbf{y} - \mathbf{D} \mathbf{c} \right\rVert_2^2
+ \lambda \left\lVert \mathbf{c} \right\rVert_1
\right\},
\end{equation}
where $\lambda \ge 0$ is a weighting factor controlling sparsity, or to use related adaptive basis-selection algorithms~\cite{jakeman2015enhancing, loukrezis2020robust, luethen2022automatic, loukrezis2025multivariate}.
In this work, sparse \glspl{pce} are computed with the hybrid \gls{lar} method~\cite{blatman2011adaptive}, which uses \gls{lar} to rank the candidate basis terms, re-estimates the coefficients of each resulting candidate basis by \gls{ols}, and selects the basis with the smallest (corrected) \gls{loo} error.
The final sparse \gls{pce} is therefore an \gls{ols} fit on a data-dependent basis.

\section{Adaptive jackknife+ conformal prediction for polynomial chaos regression}
\label{sec:acp-pce}
In the following, we present the adaptive jackknife+ \gls{cp} framework for \glspl{pce} developed in this work. 
Therefore, in this section, the \gls{cp} intervals are computed using normalized non-conformity scores as introduced in Eq.~\eqref{eq:normalized-score} and then adapted as in Eq.~\eqref{eq:loo-residuals-normalized} for jackknife \gls{cp}.
The main component of the suggested approach is the local prediction uncertainty estimator, discussed in Section~\ref{sec:acp-pce-eta}.
Subsequently, we distinguish between fixed-basis \glspl{pce} computed by means of \gls{ols} in Section~\ref{sec:acp-pce-fixed-ols}, and sparse \glspl{pce} computed with dedicated adaptive basis-selection algorithms in Section~\ref{sec:acp-pce-sparse}, as each case requires different handling.
Note that the non-adaptive counterpart of the present methodology is already available in the literature~\cite{loukrezis2026conformalized, hatstatt2026conformal} and is retrieved simply by setting the local uncertainty estimator equal to one. 

\subsection{Local prediction uncertainty estimator}
\label{sec:acp-pce-eta}
To adapt the prediction intervals to the local, input-dependent prediction uncertainty, a local uncertainty estimator $\widehat{\eta}$ as introduced in Section~\ref{sec:non-conformity-scores} is required.
To that end, an auxiliary \gls{pce} model is employed, denoted by $\widehat{\gamma}$ and given as
\begin{equation}
 \label{eq:pce-gamma}
\widehat{\gamma} \left(\mathbf{x} \right) = \sum_{k=1}^{K_\gamma} \widehat{c}_{k}^{\gamma} \Phi_{k} \left( \mathbf{x} \right),
\end{equation}
where $\widehat{c}_{k}^{\gamma}$ and $\Phi_{k}$ respectively denote the corresponding coefficients and polynomials.
This auxiliary \gls{pce} is trained on a dataset with log-transformed \gls{loo} residuals as targets, which is constructed as
\begin{equation}
\label{eq:aux-dataset}
\mathcal{D}_{\text{train}}^{\text{aux}} = \left\{ \left(\mathbf{x}^{(m)}, z_m = \ln R_m^{\text{LOO}}\right) \right\}_{m=1}^M,
\end{equation}
where $\left(R_m^{\text{LOO}}\right)_{m=1}^M$ are obtained from Eq.~\eqref{eq:loo-residuals}.
As with the base \gls{pce} model $\widehat\mu$, $\widehat\gamma$ can be fitted either by means of \gls{ols} on a fixed basis, or with a dedicated sparse basis-selection algorithm. 
This choice is made independently for $\widehat\mu$ and $\widehat\gamma$.
The local uncertainty estimator is then obtained by exponentiating the predictions of $\widehat{\gamma}$, such that
\begin{equation}
\label{eq:eta-definition}
\widehat{\eta}\left(\mathbf{x}\right) = \exp\left(\widehat{\gamma}\left(\mathbf{x}\right)\right).
\end{equation}
The log-transformation of the \gls{loo} residuals and the exponentiation in Eq.~\eqref{eq:aux-dataset} guarantees that $\widehat\eta(\mathbf{x}) > 0$, as required by Eq.~\eqref{eq:normalized-score}. 
Moreover, it stabilizes the auxiliary regression problem against heavy-tailed \gls{loo} residual magnitudes.

Two empirical stabilization measures are additionally applied to the auxiliary targets $z_m$, $m=1, \dots,M$.
First, the residuals are floored at $10^{-3}$ times their median, which keeps their logarithms finite.
Second, the log-residuals are winsorized at their empirical $5\%$ and $99\%$ quantiles, respectively denoted by $z_{\text{lo}}$ and $z_{\text{hi}}$, and the local uncertainty estimator is bounded accordingly, i.e.,
\begin{equation}
\label{eq:eta-bounded}
\widehat{\eta}\left(\mathbf{x}\right) = \min\left\{ \max\left\{ \exp\left(\widehat{\gamma}\left(\mathbf{x}\right)\right), e^{z_{\text{lo}}} \right\}, e^{z_{\text{hi}}} \right\}.
\end{equation}
The bounds are asymmetric on purpose.
The lower bound protects against near-zero residuals, which carry little information about the local uncertainty but produce excessively large normalized scores.
In contrast, the upper bound retains the large residuals that characterize the high-uncertainty regions and clips only the most extreme values. 
As a bound on $\widehat\eta$, $z_{\text{hi}}$ additionally prevents the exponential from diverging when $\widehat\gamma$ is evaluated outside the range of the training data.
In the following, the bounds of Eq.~\eqref{eq:eta-bounded} are applied to every evaluation of $\widehat\eta$ and its \gls{loo} counterparts.

As described in Section~\ref{sec:jackknife-cp}, a \gls{loo} counterpart of the local uncertainty estimator, $\widehat{\eta}_{-m}$, $m=1,\dots,M$, must be evaluated at the held-out input $\mathbf{x}^{(m)}$ and at the test input $\mathbf{x}^*$, to compute the normalized non-conformity scores in Eq.~\eqref{eq:loo-residuals-normalized} and the jackknife+ \gls{cp} interval in Eq.~\eqref{eq:jackknife-plus-interval-adaptive}.
Computing $\widehat\eta_{-m}$ is more involved than computing $\widehat\mu_{-m}$, because the training targets of $\widehat{\gamma}$ are themselves \gls{loo} quantities of $\widehat{\mu}$.
In particular, for $j \neq m$, the absolute \gls{loo} residual $R_j^{\text{LOO}}$ in Eq.~\eqref{eq:aux-dataset} is computed using the \gls{loo} model $\widehat\mu_{-j}$ which is trained on $\mathcal{D}_{-j} = \mathcal{D}_{\text{train}} \setminus \left\{ \left(\mathbf{x}^{(j)}, y^{(j)}\right) \right\}$. 
However, this dataset still contains the point $\left(\mathbf{x}^{(m)}, y^{(m)}\right)$.
Consequently, simply excluding the $j=m$ entry of $\mathcal{D}_{\text{train}}^{\text{aux}}$ when training the \gls{loo} auxiliary model $\widehat\gamma_{-m}$ does not fully remove the influence of $y^{(m)}$, which enters indirectly through every $\widehat\mu_{-j}$, $j \neq m$.

To preserve the $(1-2\alpha)$ coverage guarantee of jackknife+ \gls{cp}, each \gls{loo} quantity must be obtained from a regression algorithm applied to $\mathcal{D}_{-m}$ alone.
To that end, ``nested'' \gls{loo} residuals must be computed by double-omission of training data points, given as
\begin{equation}
\label{eq:loo-residual-nested}
R_j^{\text{LOO},-m} = \left| r_j^{\text{LOO},-m} \right| = \left| y^{(j)} - \widehat{\mu}_{-\{j,m\}}\left(\mathbf{x}^{(j)}\right) \right|, \quad j \neq m,
\end{equation}
where $\widehat\mu_{-\{j,m\}}$ is trained on $\mathcal{D}_{\text{train}} \setminus \left\{ \left(\mathbf{x}^{(j)}, y^{(j)}\right), \left(\mathbf{x}^{(m)}, y^{(m)}\right) \right\}$, i.e., excluding both points $j$ and $m$.
The \gls{loo} auxiliary model $\widehat\gamma_{-m}$ is then trained on $\left\{ \left(\mathbf{x}^{(j)}, z_j^{(-m)} = \ln R_j^{\text{LOO},-m}\right) \right\}_{j \neq m}$ and $\widehat{\eta}_{-m}\left(\mathbf{x}\right) = \exp\left(\widehat{\gamma}_{-m}\left(\mathbf{x}\right)\right)$.
Then $\widehat\mu_{-m}$ and $\widehat\eta_{-m}$ are functions of $\mathcal{D}_{-m}$ alone, since $\left(\mathbf{x}^{(m)}, y^{(m)}\right)$ does not enter the training of $\widehat\mu_{-\{j,m\}}$ or $\widehat\gamma_{-m}$, and the fitting procedure is invariant to permutations of its training data.
The aforementioned stabilization measures are applied in the same manner but
separately for each $m$, i.e., the floor and the bounds $z_{\text{lo}}^{(-m)}$, $z_{\text{hi}}^{(-m)}$ are computed from the nested residuals $\left(R_j^{\text{LOO},-m}\right)_{j \neq m}$, such that $\widehat\eta_{-m}$ remains a function of $\mathcal{D}_{-m}$ alone.
Hence, the $(1-2\alpha)$ coverage guarantee of jackknife+ \gls{cp} extends to the adaptive prediction intervals as well. 

Preserving the coverage guarantee requires the computation of $M(M-1)$ nested \gls{loo} residuals, obtained from $M(M-1)/2$ distinct double-omission models $\widehat\mu_{-\{j,m\}}$ (since $\widehat\mu_{-\{j,m\}} = \widehat\mu_{-\{m,j\}}$).
Together with the $M$ \gls{loo} base models $\widehat\mu_{-m}$, the number of base models to be trained is $M(M+1)/2$, while $M$ \gls{loo} auxiliary models $\widehat\gamma_{-m}$ must be trained as well.
If all these models were to be trained explicitly, the overall computational cost would quickly become prohibitive.
Section~\ref{sec:acp-pce-fixed-ols} shows that repeated model training can be bypassed altogether for fixed-basis, \gls{ols}-fitted \glspl{pce}, which admit analytic, closed-form expressions for all necessary quantities.
As these closed-form expressions are not applicable for sparse \glspl{pce}, Section~\ref{sec:acp-pce-sparse} introduces a heuristic simplification to mitigate the computational demand.

\subsection{Adaptive jackknife+ CP for fixed-basis OLS-fitted PCEs}
\label{sec:acp-pce-fixed-ols}
We first consider fixed-basis \glspl{pce} computed by solving the \gls{ols} problem in Eq.~\eqref{eq:regression-matrix}. 
In this case, one can leverage the linear-in-parameters structure of the resulting \gls{pce} models (base model $\widehat{\mu}$ and auxiliary model $\widehat{\gamma}$) to avoid the aforementioned repeated model training, thus significantly accelerating the estimation of the jackknife+ \gls{cp} intervals.
In particular, all quantities required by the guarantee-preserving method presented in Section~\ref{sec:acp-pce-eta} can be computed in closed form and at no additional model training cost, from a single \gls{ols} fit of $\widehat\mu$ and without explicitly training any auxiliary model. 

Following the notation introduced in Section~\ref{sec:pce}, we denote the rows of the design matrix of $\widehat\mu$, $ \mathbf{D} \in \mathbb{R}^{M \times K}$, by $\boldsymbol{\psi}_m^\top = \left(\Psi_1\left(\mathbf{x}^{(m)}\right), \dots, \Psi_K\left(\mathbf{x}^{(m)}\right)\right)$, $m = 1, \dots, M$. 
A so-called hat matrix is then formed as $\mathbf{H} = \mathbf{D}\left(\mathbf{D}^\top\mathbf{D}\right)^{-1}\mathbf{D}^\top$, with entries $h_{jm} = \boldsymbol{\psi}_j^\top \left(\mathbf{D}^\top\mathbf{D}\right)^{-1} \boldsymbol{\psi}_m$.
Moreover, let $e_m = y^{(m)} - \widehat\mu\left(\mathbf{x}^{(m)}\right)$ denote the residuals of the base model $\widehat\mu$ trained on the full training dataset $\mathcal{D}_{\text{train}}$.
Recall from Eq.~\eqref{eq:regression-solution} that the coefficient vector of $\widehat\mu$ is
$\widehat{\mathbf{c}} = \left(\mathbf{D}^\top\mathbf{D}\right)^{-1}\mathbf{D}^\top\mathbf{y}$.
Removing the $m$-th data point results in a rank-$1$ downdate of $\mathbf{D}^\top\mathbf{D}$. 
Using the Sherman--Morrison--Woodbury formula~\cite{seber2003linear}, the coefficients of the \gls{loo} model $\widehat\mu_{-m}$ are given by
\begin{equation}
\label{eq:loo-coefficients-analytical}
\widehat{\mathbf{c}}_{-m} = \widehat{\mathbf{c}} - \left(\mathbf{D}^\top\mathbf{D}\right)^{-1}
\boldsymbol{\psi}_m \, \frac{e_m}{1 - h_{mm}}.
\end{equation}
Evaluating $\widehat\mu_{-m}(\mathbf{x}) = \boldsymbol{\psi}(\mathbf{x})^\top \widehat{\mathbf{c}}_{-m}$ at the omitted input $\mathbf{x}^{(m)}$ yields the signed \gls{loo} residual in Eq.~\eqref{eq:loo-residuals},
\begin{equation}
\label{eq:loo-residual-analytical}
r_m^{\text{LOO}} = y^{(m)} - \boldsymbol{\psi}_m^\top \mathbf{c}_{-m}
= \frac{e_m}{1 - h_{mm}},
\end{equation}
and evaluating it at a test input $\mathbf{x}^*$ yields
\begin{equation}
\label{eq:loo-model-evaluation-analytical}
\widehat{\mu}_{-m}\left(\mathbf{x}^*\right) = \boldsymbol{\psi}_*^\top
\widehat{\mathbf{c}}_{-m} = \widehat{\mu}\left(\mathbf{x}^*\right) -
\boldsymbol{\psi}_*^\top \left(\mathbf{D}^\top \mathbf{D}\right)^{-1}
\boldsymbol{\psi}_m \, r_m^{\text{LOO}},
\end{equation}
where $\boldsymbol{\psi}_*^\top = \left(\Psi_1\left(\mathbf{x}^*\right), \dots,
\Psi_K\left(\mathbf{x}^*\right)\right)$.
Note that the analytic expressions given in Eqs.~\eqref{eq:loo-residual-analytical} and \eqref{eq:loo-model-evaluation-analytical} suffice to compute the non-adaptive jackknife+ \gls{cp} interval in Eq.~\eqref{eq:jackknife-plus-interval-nonadaptive}~\cite{loukrezis2026conformalized, hatstatt2026conformal}.

The nested \gls{loo} residuals defined in Eq.~\eqref{eq:loo-residual-nested} admit an analogous closed form.
Removing two data points $j \neq m$ from the training dataset results in a rank-$2$ downdate of $\mathbf{D}^\top\mathbf{D}$.
Again applying the Sherman--Morrison--Woodbury formula, the coefficients of the double-omission model $\widehat\mu_{-\{j,m\}}$ are given by
\begin{equation}
\label{eq:nested-coefficients-analytical}
\widehat{\mathbf{c}}_{-\{j,m\}} = \widehat{\mathbf{c}} - \left(\mathbf{D}^\top\mathbf{D}\right)^{-1} \begin{bmatrix} \boldsymbol{\psi}_j & \boldsymbol{\psi}_m \end{bmatrix} \left(\mathbf{I}_2 - \mathbf{H}_{\{j,m\}}\right)^{-1} \begin{bmatrix} e_j \\ e_m \end{bmatrix},
\quad
\mathbf{H}_{\{j,m\}} = \begin{bmatrix} h_{jj} & h_{jm} \\ h_{jm} & h_{mm} \end{bmatrix},
\end{equation}
which is the two-point counterpart of Eq.~\eqref{eq:loo-coefficients-analytical}. 
Here, $\mathbf{I}_2$ is the $2 \times 2$ identity matrix.
Evaluating $\widehat\mu_{-\{j,m\}}$ at the two omitted inputs yields both nested \gls{loo} residuals associated with the pair $\{j, m\}$ at once, i.e., 
\begin{equation}
\label{eq:loo-residual-nested-analytical}
\begin{bmatrix} r_j^{\text{LOO},-m} \\ r_m^{\text{LOO},-j} \end{bmatrix}
= \left(\mathbf{I}_2 - \mathbf{H}_{\{j,m\}}\right)^{-1} \begin{bmatrix} e_j \\ e_m \end{bmatrix}
= \frac{1}{\Delta_{jm}} \begin{bmatrix} \left(1 - h_{mm}\right) e_j + h_{jm} e_m \\ h_{jm} e_j + \left(1 - h_{jj}\right) e_m \end{bmatrix},
\end{equation}
with $\Delta_{jm} = \left(1 - h_{jj}\right)\left(1 - h_{mm}\right) - h_{jm}^2$.
This is the two-point equivalent of Eq.~\eqref{eq:loo-residual-analytical}, to which it reduces for $h_{jm} = 0$.
In contrast to Eq.~\eqref{eq:loo-residual-analytical}, which only requires the diagonal elements of the hat matrix, the two-point expression additionally requires the off-diagonal entries $h_{jm}$.

For $m=1,\dots,M$, the \gls{loo} auxiliary model $\widehat\gamma_{-m}$ is trained on the targets $z_j^{(-m)} = \ln R_j^{\text{LOO},-m}$, $j \neq m$.
Let $\mathbf{D}_\gamma \in \mathbb{R}^{M \times K_\gamma}$ denote the design matrix of $\widehat\gamma$, with rows $\boldsymbol{\phi}_m^\top = \left(\Phi_1\left(\mathbf{x}^{(m)}\right), \dots, \Phi_{K_\gamma}\left(\mathbf{x}^{(m)}\right)\right)$.
The diagonal entries of the corresponding hat matrix $\mathbf{H}_{\gamma}$ are given by  $h_{mm}^{\gamma} = \boldsymbol{\phi}_m^\top \left(\mathbf{D}_\gamma^\top\mathbf{D}_\gamma\right)^{-1} \boldsymbol{\phi}_m$.
Since the targets $z_j^{(-m)}$ change with $m$ for every $j$, the single-point identity in Eq.~\eqref{eq:loo-residual-analytical} cannot be applied to $\widehat\gamma$ directly.
However, as only a single row is removed from $\mathbf{D}_\gamma$, the Sherman--Morrison--Woodbury formula still applies to the corresponding normal equations. 
Hence, the coefficients of $\widehat\gamma_{-m}$ are obtained as 
\begin{equation}
\label{eq:aux-loo-coefficients-analytical}
\widehat{\mathbf{c}}^{\gamma}_{-m} = \left[ \left(\mathbf{D}_\gamma^\top\mathbf{D}_\gamma\right)^{-1} + \frac{\left(\mathbf{D}_\gamma^\top\mathbf{D}_\gamma\right)^{-1} \boldsymbol{\phi}_m \boldsymbol{\phi}_m^\top \left(\mathbf{D}_\gamma^\top\mathbf{D}_\gamma\right)^{-1}}{1 - h_{mm}^{\gamma}} \right] \sum_{j \neq m} \boldsymbol{\phi}_j \, z_j^{(-m)}.
\end{equation}
The \gls{loo} local uncertainty estimator is then evaluated at any input $\mathbf{x}$ as
\begin{equation}
\label{eq:eta-loo-analytical}
\widehat\eta_{-m}\left(\mathbf{x}\right) = \exp\left( \boldsymbol{\phi}\left(\mathbf{x}\right)^\top \widehat{\mathbf{c}}^{\gamma}_{-m} \right),
\end{equation}
with $\boldsymbol{\phi}(\mathbf{x})^\top = \left(\Phi_1(\mathbf{x}), \dots, \Phi_{K_\gamma}(\mathbf{x})\right)$, in particular at the held-out input $\mathbf{x}^{(m)}$ and at the test input $\mathbf{x}^*$.
Note that the auxiliary model $\widehat\gamma$ is never explicitly trained on the full auxiliary dataset $\mathcal{D}_{\text{train}}^{\text{aux}}$, as only the factorization of $\mathbf{D}_\gamma^\top\mathbf{D}_\gamma$ is needed.
Also note that the expressions above require the corresponding \gls{ols} problems to remain well-posed after the omission of the data points, i.e., that $\mathbf{D}$ retains full column rank after the removal of any two rows (which requires $M \geq K + 2$), and that $\mathbf{D}_\gamma$ retains full column rank after the removal of any single row.

Using Eqs.~\eqref{eq:loo-residual-analytical} and \eqref{eq:eta-loo-analytical}, the normalized non-conformity scores are computed as in Eq.~\eqref{eq:loo-residuals-normalized}. 
Consequently, the adaptive jackknife+ \gls{cp} interval is constructed as in Eq.~\eqref{eq:jackknife-plus-interval-adaptive}, where the \gls{loo} model evaluation $\widehat\mu_{-m}\left(\mathbf{x}^*\right)$ at the test point is obtained from Eq.~\eqref{eq:loo-model-evaluation-analytical}.
The $(1-2\alpha)$ marginal coverage guarantee of jackknife+ \gls{cp} is accordingly retained.

Regarding the computational cost, $\mathbf{D}^\top\mathbf{D}$ and $\mathbf{D}_\gamma^\top\mathbf{D}_\gamma$ are factorized only once, typically via thin singular value decompositions of $\mathbf{D}$ and $\mathbf{D}_\gamma$, at a cost of $\mathcal{O}\left(M K^2\right)$ and $\mathcal{O}\left(M K_\gamma^2\right)$ operations, respectively.
The quantities that do not depend on the test input are then obtained as follows.
Assembling the full hat matrix $\mathbf{H}$ requires $\mathcal{O}\left(M^2 K\right)$ operations, and the $M(M-1)$ nested residuals in Eq.~\eqref{eq:loo-residual-nested-analytical}, together with their logarithms, require $\mathcal{O}\left(M^2\right)$ operations.
The $M$ coefficient vectors in Eq.~\eqref{eq:aux-loo-coefficients-analytical} require $\mathcal{O}\left(M^2 K_\gamma + M K_\gamma^2\right)$ operations in total.
The \gls{loo} residuals from Eq.~\eqref{eq:loo-residual-analytical}, the estimates $\widehat\eta_{-m}\left(\mathbf{x}^{(m)}\right)$, and the normalized scores from Eq.~\eqref{eq:loo-residuals-normalized} then follow with $\mathcal{O}\left(M \left(K + K_\gamma\right)\right)$ operations.
For each test input $\mathbf{x}^*$, the $M$ \gls{loo} model evaluations in Eq.~\eqref{eq:loo-model-evaluation-analytical} require $\mathcal{O}\left(K^2 + M K\right)$ operations, since $\boldsymbol{\psi}_*^\top \left(\mathbf{D}^\top\mathbf{D}\right)^{-1}$ is computed once and then multiplied by $\mathbf{D}^\top$.
The $M$ \gls{loo} local uncertainty estimates $\widehat\eta_{-m}\left(\mathbf{x}^*\right)$ in Eq.~\eqref{eq:eta-loo-analytical} require $\mathcal{O}\left(M K_\gamma\right)$ operations. 
The two quantiles that define the interval endpoints are obtained in $\mathcal{O}\left(M\right)$ operations by linear-time selection.
For $L$ test inputs, the total cost thus amounts to
\begin{equation}
\label{eq:acp-cost}
\mathcal{O}\Big( \underbrace{M^2 \left(K + K_\gamma\right) + M \left(K^2 + K_\gamma^2\right)}_{\text{independent of the test inputs}} + \underbrace{L \left[ K^2 + M \left(K + K_\gamma\right) \right]}_{\text{test-input evaluations}} \Big)
\end{equation}
operations, with $\mathcal{O}\left(M^2\right)$ memory requirement for $\mathbf{H}$ and the nested residuals.
The test-input component is inherent to jackknife+ \gls{cp}, adaptive or not.
The component that is independent of the test inputs scales quadratically with $M$, but consists solely of (dense) matrix products.
In contrast, explicitly training the $M(M+1)/2$ base models $\widehat\mu_{-m}$ and $\widehat\mu_{-\{j,m\}}$ and the $M$ auxiliary models $\widehat\gamma_{-m}$ would require $\mathcal{O}\left(M^3 K^2 + M^2 K_\gamma^2\right)$ operations.

\subsection{Adaptive jackknife+ CP for sparse PCEs}
\label{sec:acp-pce-sparse}
The case of sparse \glspl{pce} computed with adaptive basis-selection algorithms deserves special attention, as the closed-form expressions of Section~\ref{sec:acp-pce-fixed-ols} are not directly applicable.
For a given sparse basis, a subsequent \gls{ols} fit still admits these expressions.
However, the basis selection procedure depends on the available training data.
Naively applying the closed-form expressions to a sparse \gls{pce}, for example, by ``freezing'' the basis selected on the full training dataset, lets the omitted data points influence the \gls{loo} models through the basis selection.
This can lead to a severe degradation in empirical coverage, especially in small-data settings where the basis selection is highly sensitive to the available training data~\cite{hatstatt2026conformal}.
Therefore, whenever a sparse \gls{pce} is employed, its \gls{loo} counterparts must be trained explicitly, each performing its own basis selection from the same candidate basis on the corresponding reduced dataset.
We distinguish two cases, depending on whether the base model $\widehat\mu$ or the auxiliary model $\widehat\gamma$ is a sparse \gls{pce}.

If $\widehat\mu$ is a fixed-basis, \gls{ols}-fitted \gls{pce} and only $\widehat\gamma$ is a sparse \gls{pce}, the \gls{loo} residuals, the \gls{loo} model evaluations, and the nested \gls{loo} residuals remain available in closed form via Eqs.~\eqref{eq:loo-residual-analytical}, \eqref{eq:loo-model-evaluation-analytical}, and \eqref{eq:loo-residual-nested-analytical}, respectively.
Only the $M$ \gls{loo} auxiliary models $\widehat\gamma_{-m}$ must be trained explicitly, each on $\left\{ \left(\mathbf{x}^{(j)}, \ln R_j^{\text{LOO},-m}\right) \right\}_{j \neq m}$, and evaluated at $\mathbf{x}^{(m)}$ and $\mathbf{x}^*$.
In this case, the guarantee-preserving method of Section~\ref{sec:acp-pce-eta} can be applied at the additional cost of $M$ sparse auxiliary model fits, and no further simplification is required.

If $\widehat\mu$ is a sparse \gls{pce}, the $M$ \gls{loo} models $\widehat\mu_{-m}$ must be trained explicitly on $\mathcal{D}_{-m}$, $m = 1, \dots, M$, to provide the \gls{loo} residuals $R_m^{\text{LOO}}$ and the \gls{loo} model evaluations $\widehat\mu_{-m}\left(\mathbf{x}^*\right)$.
The nested \gls{loo} residuals in Eq.~\eqref{eq:loo-residual-nested}, however, would require the double-omission models $\widehat\mu_{-\{j,m\}}$ as well, i.e., $M(M+1)/2$ sparse fits of the base model in total, which quickly becomes prohibitive.
We therefore resort to a heuristic simplification, which forgoes the formal $(1-2\alpha)$ coverage guarantee of jackknife+ \gls{cp}.
In particular, the readily available \gls{loo} residuals $R_j^{\text{LOO}}$ are reused as training targets of the \gls{loo} auxiliary models (after their log-transform), such that $\widehat\gamma_{-m}$ is trained on $\mathcal{D}_{-m}^{\text{aux}} = \mathcal{D}_{\text{train}}^{\text{aux}} \setminus \left\{ \left(\mathbf{x}^{(m)}, z_m = \ln R_m^{\text{LOO}}\right) \right\}$, with $\mathcal{D}_{\text{train}}^{\text{aux}}$ given as in Eq.~\eqref{eq:aux-dataset}.
As discussed in Section~\ref{sec:acp-pce-eta}, $y^{(m)}$ now affects indirectly the computation of $\widehat\gamma_{-m}$  through the \gls{loo} models $\widehat\mu_{-j}$, $j \neq m$.
Under this simplification, all \gls{loo} auxiliary models share the same targets $z_j = \ln R_j^{\text{LOO}}$, and $\widehat\gamma_{-m}$ differs from $\widehat\gamma$ only by the omission of the $m$-th row of $\mathcal{D}_{\text{train}}^{\text{aux}}$, whereas in the guarantee-preserving method each $\widehat\gamma_{-m}$ has its own targets $z_j^{(-m)} = \ln R_j^{\text{LOO},-m}$.
Hence, the dependence of $z_j$ on $y^{(m)}$ through $\widehat\mu_{-j}$ is ignored and the coverage guarantee is lost. 
Since only a single row is removed, if $\widehat\gamma$ is a fixed-basis \gls{ols}-fitted \gls{pce}, the single-point identities
in Eqs.~\eqref{eq:loo-coefficients-analytical}, \eqref{eq:loo-residual-analytical}, and \eqref{eq:loo-model-evaluation-analytical} apply directly to the auxiliary model.
If $\widehat\gamma$ is sparse as well, the $M$ \gls{loo} auxiliary models $\widehat\gamma_{-m}$ are trained explicitly on $\mathcal{D}_{-m}^{\text{aux}}$ and evaluated at $\mathbf{x}^{(m)}$ and $\mathbf{x}^*$.
In either case, $\widehat\eta_{-m}\left(\mathbf{x}^{(m)}\right) = \exp\left(\widehat\gamma_{-m}\left(\mathbf{x}^{(m)}\right)\right)$ is used to compute the normalized non-conformity scores $s_m^{\text{n}}$ as in Eq.~\eqref{eq:loo-residuals-normalized}.

Algorithmically, the adaptive jackknife+ \gls{cp} interval for a sparse base \gls{pce} model $\widehat\mu$ is obtained as follows.
\begin{enumerate}
    \item For $m = 1, \dots, M$, train the \gls{loo} model $\widehat\mu_{-m}$ on $\mathcal{D}_{-m}$, including its own basis selection, and compute the absolute \gls{loo} residual $R_m^{\text{LOO}}$ and the \gls{loo} model evaluation $\widehat\mu_{-m}\left(\mathbf{x}^*\right)$.
    \item Form the auxiliary dataset $\mathcal{D}_{\text{train}}^{\text{aux}}$ as in Eq.~\eqref{eq:aux-dataset}. 
    If $\widehat\gamma$ is a fixed-basis OLS-fitted \gls{pce}, fit $\widehat\gamma$ on $\mathcal{D}_{\text{train}}^{\text{aux}}$.
    \item For $m = 1, \dots, M$, obtain $\widehat\gamma_{-m}\left(\mathbf{x}^{(m)}\right)$, either in closed form if $\widehat\gamma$ is a fixed-basis \gls{ols}-fitted \gls{pce}, or by explicitly training $M$ \gls{loo} auxiliary models on $\mathcal{D}_{-m}^{\text{aux}}$ if $\widehat\gamma$ is a sparse \gls{pce}.
    \item For $m = 1, \dots, M$, compute $\widehat\eta_{-m}\left(\mathbf{x}^{(m)}\right) = \exp\left(\widehat\gamma_{-m}\left(\mathbf{x}^{(m)}\right)\right)$ and the normalized non-conformity scores $s_m^{\text{n}}$ as in Eq.~\eqref{eq:loo-residuals-normalized}.
    \item Compute the adaptive jackknife+ \gls{cp} intervals as in Eq.~\eqref{eq:jackknife-plus-interval-adaptive}.
\end{enumerate}
In total, this requires $M$ sparse fits of the base model and, if $\widehat\gamma$ is sparse, $M$ sparse fits of the auxiliary model. 
This is a substantial reduction compared to the $M\left(M+1\right)/2$ sparse base models plus $M$ sparse auxiliary models that would be explicitly trained in the guarantee-preserving method.
While this heuristic approach can still be computationally demanding in general, sparse \glspl{pce} are commonly employed in small-data settings, where the overall computational demand remains manageable.
Although the formal coverage guarantee is lost, the empirical coverage is typically on par with the target level, as shown in the numerical experiments of Section~\ref{sec:num-exp}.

\section{Numerical experiments}
\label{sec:num-exp}
In the following, we apply and evaluate the suggested adaptive jackknife+ \gls{cp}-\gls{pce} method on different regression problems, first using a benchmark model and subsequently an engineering simulation dataset.
The following evaluation metrics are employed, computed on a test dataset $\mathcal{D}_{\mathrm{test}} = \left\{\left(\mathbf{x}^{(l)}, y^{(l)}\right)\right\}_{l=1}^L$ that is separate from the training and, if relevant, calibration datasets.
\begin{description}
\item[Empirical coverage:] Empirical coverage is estimated by computing the fraction of test points whose true output lies within the prediction interval, and compared against the prescribed target coverage level.
\item[Mean interval width:] The statistical efficiency of the \gls{cp} intervals is assessed by their mean width over the test dataset, where narrower intervals are preferable for comparable coverage.
\item[Winkler score:] Let  $\left\{ \left[\ell^{(l)}, u^{(l)}\right] \right\}_{l=1}^L$ be the prediction intervals corresponding to the test data. 
Then, the so-called Winkler score~\cite{gneiting2007strictly} is computed as
\begin{equation}
\label{eq:winkler}
W = \frac{1}{L} \sum_{l=1}^L \left[ \left(u^{(l)} - \ell^{(l)} \right) + \frac{2}{\alpha}\left(\ell^{(l)} - y^{(l)}\right)_+ + \frac{2}{\alpha}\left(y^{(l)} - u^{(l)}\right)_+ \right],
\end{equation}
where $\left(z\right)_+ = \max\left(0, z\right)$. 
This scoring rule rewards narrow intervals (first term in Eq.~\eqref{eq:winkler}), while simultaneously penalizing observations that fall outside the interval (second and third terms in Eq.~\eqref{eq:winkler}).
In contrast to the interval width alone, the Winkler score also reflects how well the interval widths are distributed.
It follows that a lower score indicates a better prediction interval.
\end{description}
The accuracy of the base \gls{pce} model $\widehat{\mu}$ is reported in terms of the coefficient of determination on the test dataset:
\begin{equation}
R_{\text{test}}^2 = 1 - \frac{\sum_{l=1}^{L} \left( y^{(l)} - \widehat{\mu}\left(\mathbf{x}^{(l)}\right) \right)^2}{\sum_{l=1}^{L} \left( y^{(l)} - \bar{y} \right)^2},
\quad
\bar{y} = \frac{1}{L} \sum_{l=1}^{L} y^{(l)}.
\end{equation}
To assess the variability of all metrics, all numerical experiments are repeated multiple times ($N_{\mathrm{rep}}=100$ repetitions are used throughout), by either generating new  datasets (benchmarking use case, Section~\ref{sec:benchmark}) or by randomly reshuffling the original dataset before its partitioning (engineering use case, Section~\ref{sec:thyc}).

The adaptive jackknife+ \gls{cp}-\gls{pce} method developed in this work is compared against its non-adaptive counterpart~\cite{loukrezis2026conformalized, hatstatt2026conformal}, as well as against adaptive and non-adaptive split \gls{cp}-\glspl{pce}.
When adaptive split \gls{cp} is employed, the local uncertainty estimator is constructed as in Section~\ref{sec:acp-pce-eta}, except that the auxiliary model $\widehat\gamma$ is trained on the log-transformed five-fold \gls{cv} residuals~\cite{papadopoulos2011reliable}.
In this case, $\widehat\eta$ is independent of the calibration set and the finite-sample coverage guarantee of split \gls{cp} is retained.
The following abbreviations are used for brevity: (i) ``J+'' for non-adaptive jackknife+ \gls{cp}; (ii) ``AJ+'' for adaptive jackknife+ \gls{cp}; (iii) ``S'' for non-adaptive split \gls{cp}; (iv) ``AS'' for adaptive split \gls{cp}. 

All \gls{cp}-\gls{pce} variants and numerical experiments have been implemented in the Python programming language, in particular with the OpenTURNS library~\cite{baudin2017openturns}.
All computations have been performed on a standard laptop with an Intel\textsuperscript{\tiny{\circledR}} Core\textsuperscript{\tiny{\texttrademark}} Ultra 7 Processor 155H (1400~MHz, 16 cores, 22 logical processors) and 32~GB RAM.

\subsection{Benchmarking use case: Cantilever beam model}
\label{sec:benchmark}
As a benchmark model with known heteroscedastic noise, we consider a cantilever beam of length $b$, with a rectangular cross-section of width $w$ and thickness $t$, and made of a material with Young's modulus $E$.
The beam is subjected to a uniformly distributed load $F_{\mathrm{d}}$ and to an end load $F_{\mathrm{e}}$.
According to Euler--Bernoulli beam theory, the tip deflection of the beam is given by
\begin{equation}
\label{eq:cantilever}
\delta = \frac{F_{\mathrm{e}} b^3}{3 E I} + \frac{F_{\mathrm{d}} b^4}{8 E I},
\quad
I = \frac{w t^3}{12}.
\end{equation}
The model's input parameters are assumed to be mutually independent random variables following the distributions listed in Table~\ref{tab:cantilever-inputs}.
The observed random inputs are $\mathbf{X} = (b, w, t, E, F_{\mathrm{d}})$, whereas the end load $F_{\mathrm{e}}$ is also random but not observed, thus acting as a noise source.
Consequently, the tip deflection is random for a fixed input realization $\mathbf{x} = \mathbf{X}(\omega)$ and the cantilever beam model acts as a stochastic simulator.

\begin{table}[b!]
\centering
\caption{Parameters of the cantilever beam model and their probability distributions. $\mathcal{U}(\ell, u)$ denotes a uniform distribution on $[\ell, u]$. $\mathcal{LN}(m, v)$ denotes a lognormal distribution with mean $m$ and coefficient of variation $v$. The end load $F_{\mathrm{e}}$ is not observed and therefore not an input of the \gls{pce} models.}
\label{tab:cantilever-inputs}
\begin{tabular}{lll}
\hline
Parameter & Description & Distribution \\
\hline
$b$ & Beam length (m) & $\mathcal{U}(1.0, 2.0)$ \\
$w$ & Cross-section width (m) & $\mathcal{U}(0.05, 0.10)$ \\
$t$ & Cross-section thickness (m) & $\mathcal{U}(0.08, 0.12)$ \\
$E$ & Young's modulus (GPa) & $\mathcal{LN}(210, 0.05)$ \\
$F_{\mathrm{d}}$ & Distributed load (kN/m) & $\mathcal{U}(2.0, 6.0)$ \\
$F_{\mathrm{e}}$ & End load (kN) & $\mathcal{LN}(5.0, 0.25)$ \\
\hline
\end{tabular}
\end{table}

The tip deflection formula can be equivalently written as
\begin{equation}
\label{eq:cantilever-compliance}
\delta = \chi(\mathbf{x}) \left( F_{\mathrm{e}} + \frac{3}{8} F_{\mathrm{d}} b \right),
\quad
\chi(\mathbf{x}) = \frac{4 b^3}{E w t^3},
\end{equation}
where $\chi(\mathbf{x})$ is the compliance of the beam, i.e., its tip deflection per unit end load.
Since $F_{\mathrm{e}}$ is independent of $\mathbf{X}$, the tip deflection can be decomposed into a deterministic part and a zero-mean noise term. Denoting the mean of $F_{\mathrm{e}}$ by $m_{F_{\mathrm{e}}}$, Eq.~\eqref{eq:cantilever-compliance} can be rewritten as
\begin{equation}
\label{eq:cantilever-noise}
\delta =  \chi(\mathbf{x}) \left( m_{F_{\mathrm{e}}} + \frac{3}{8} F_{\mathrm{d}} b \right) + \varepsilon(\mathbf{x}), 
\qquad
\varepsilon(\mathbf{x}) = \chi(\mathbf{x}) \left( F_{\mathrm{e}} - m_{F_{\mathrm{e}}} \right),
\end{equation}
where $\varepsilon(\mathbf{x})$ is a zero-mean random variable, following a right-skewed distribution scaled by the compliance.

In this benchmark study, all \gls{cp}-\gls{pce} variants use the same $M$ training points and therefore share the same base \gls{pce}.
The split \gls{cp} variants are calibrated on an additional independent dataset of very large size ($J=10^5$), such that their results may serve as a reference.
The size of the test dataset is set to $L=10^3$ throughout.
Three significance levels are tested, i.e., $\alpha \in \left\{0.1, 0.05, 0.01\right\}$.
To assess the variability of all evaluation metrics, all numerical experiments are repeated $N_\mathrm{rep} = 100$ times. 
In each repetition, new training, calibration, and test datasets are generated by Monte Carlo sampling from the distributions given in Table~\ref{tab:cantilever-inputs}.

\subsubsection{Total-degree OLS-fitted CP-PCE}
\label{sec:cantilever-td-pce}
We first consider a training dataset of size $M=400$. 
Both the base and the auxiliary models are total-degree \gls{ols}-fitted \glspl{pce}, where $P_{\mathrm{base}}=3$ and $P_{\mathrm{aux}}=2$.
The AJ+ \gls{cp} intervals are computed with the guarantee-preserving method of Section~\ref{sec:acp-pce-eta}, using the closed-form expressions of Section~\ref{sec:acp-pce-fixed-ols}.
Hence,  all \gls{cp}-\gls{pce} variants are computed very fast: J+ in $0.016$~s, AJ+ in $0.104$~s, S in $0.356$~s, and AS in $0.685$~s (on average).
The longer computation times of the split \gls{cp} variants are attributed to the evaluation of the base and auxiliary \glspl{pce} on the $J=10^5$ calibration points.
On the test dataset, the base model yields a coefficient of determination $R_{\text{test}}^2 \approx 0.92$.

\begin{figure}[t!]
    \centering
    \begin{subfigure}{0.7\textwidth}
        \centering
        \includegraphics[width=\linewidth]{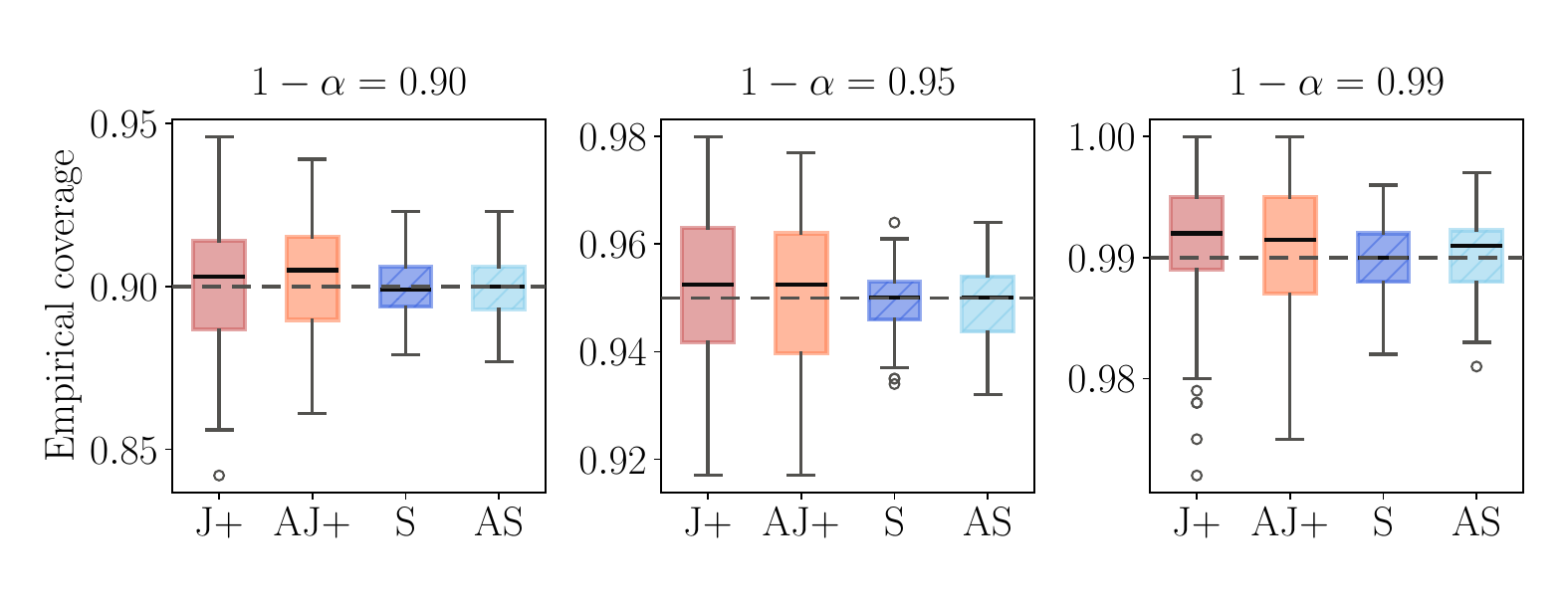}
        \caption{Empirical coverage.}
        \label{fig:cantilever-coverage}
    \end{subfigure}
    \begin{subfigure}{0.7\textwidth}
        \centering
        \includegraphics[width=\linewidth]{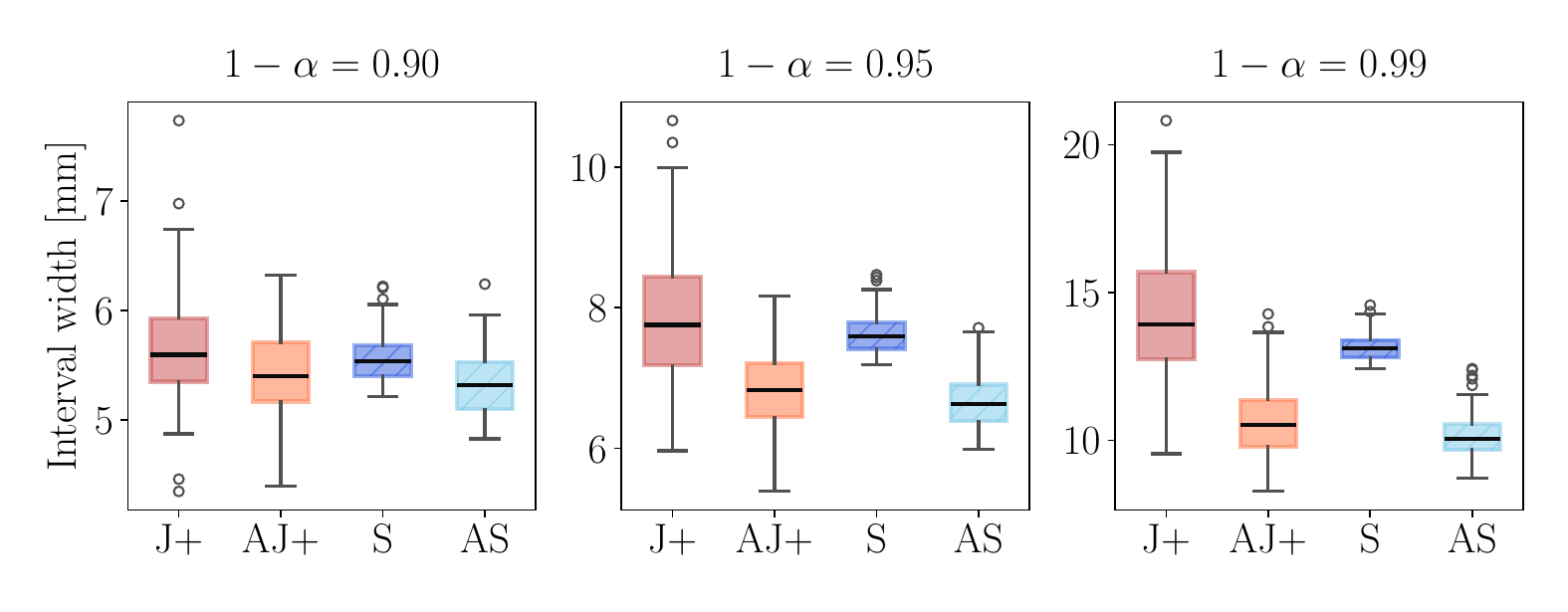}
        \caption{Mean interval width.}
        \label{fig:cantilever-intervals}
    \end{subfigure}
    \begin{subfigure}{0.7\textwidth}
        \centering
        \includegraphics[width=\linewidth]{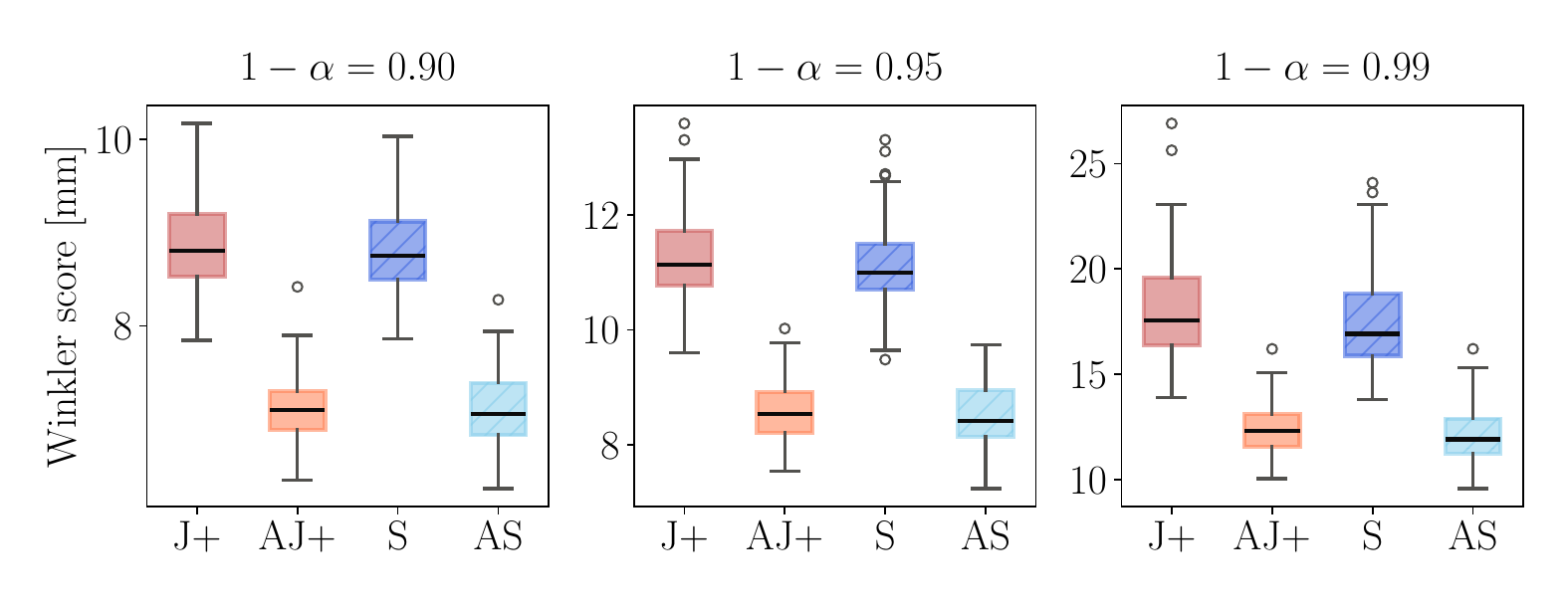}
        \caption{Winkler score.}
        \label{fig:cantilever-winkler}
    \end{subfigure}
    \caption{Box-plots of empirical coverage, mean interval width, and Winkler score by the different total-degree \gls{ols}-fitted \gls{cp}-\gls{pce} variants applied to the cantilever beam model.}
    \label{fig:cantilever-box-plots}
\end{figure}

\begin{figure}[t!]
\centering
\includegraphics[width=0.7\linewidth]{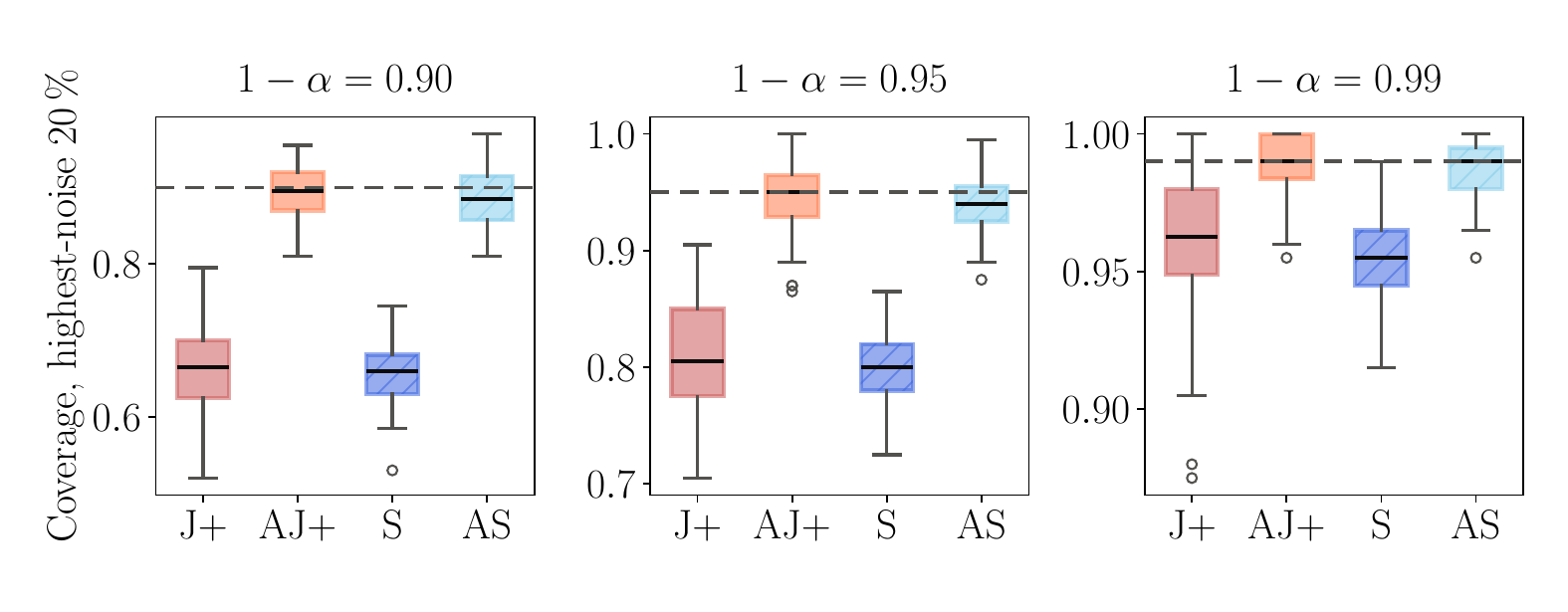}
\caption{Box-plots of the empirical coverage of the different total-degree \gls{ols}-fitted \gls{cp}-\gls{pce} variants on the 20\% of test points with the highest noise level, for the cantilever beam benchmark.}
\label{fig:cantilever-cov-high}
\end{figure}

Figure~\ref{fig:cantilever-box-plots} shows the empirical coverage, mean interval width, and Winkler score over the $N_{\mathrm{rep}}=100$ repetitions in the form of box-plots.
All \gls{cp}-\gls{pce} variants attain the target coverage, on average. 
Compared to their non-adaptive counterparts, the adaptive variants reduce the mean interval width by approximately 4\%, 13\%, and 25\%, and the Winkler score by approximately 20\%, 24\%, and 31\%, for $\alpha = 0.1, 0.05, 0.01$, respectively.
The improvement in the Winkler score exceeds the reduction in mean interval width because the adaptive intervals redistribute their width according to the noise level.
For instance, on the 20\% of test points with the highest noise level, the coverage of J+ and S drops substantially, whereas AJ+ and AS retain coverages very close to the target levels.
These results are shown in Figure~\ref{fig:cantilever-cov-high}.

The jackknife+ variants perform almost identically to their split counterparts on average, however with a larger variation in their results, especially in coverage and mean interval width.
Since all variants share the same base \gls{pce}, this difference in variation stems from the calibration step alone and is to be expected, as the split variants are calibrated on $J=10^5$ data points, whereas the jackknife+ variants are calibrated on the $M=400$ training data points only.

\begin{figure}[t!]
\centering
\includegraphics[width=\linewidth]{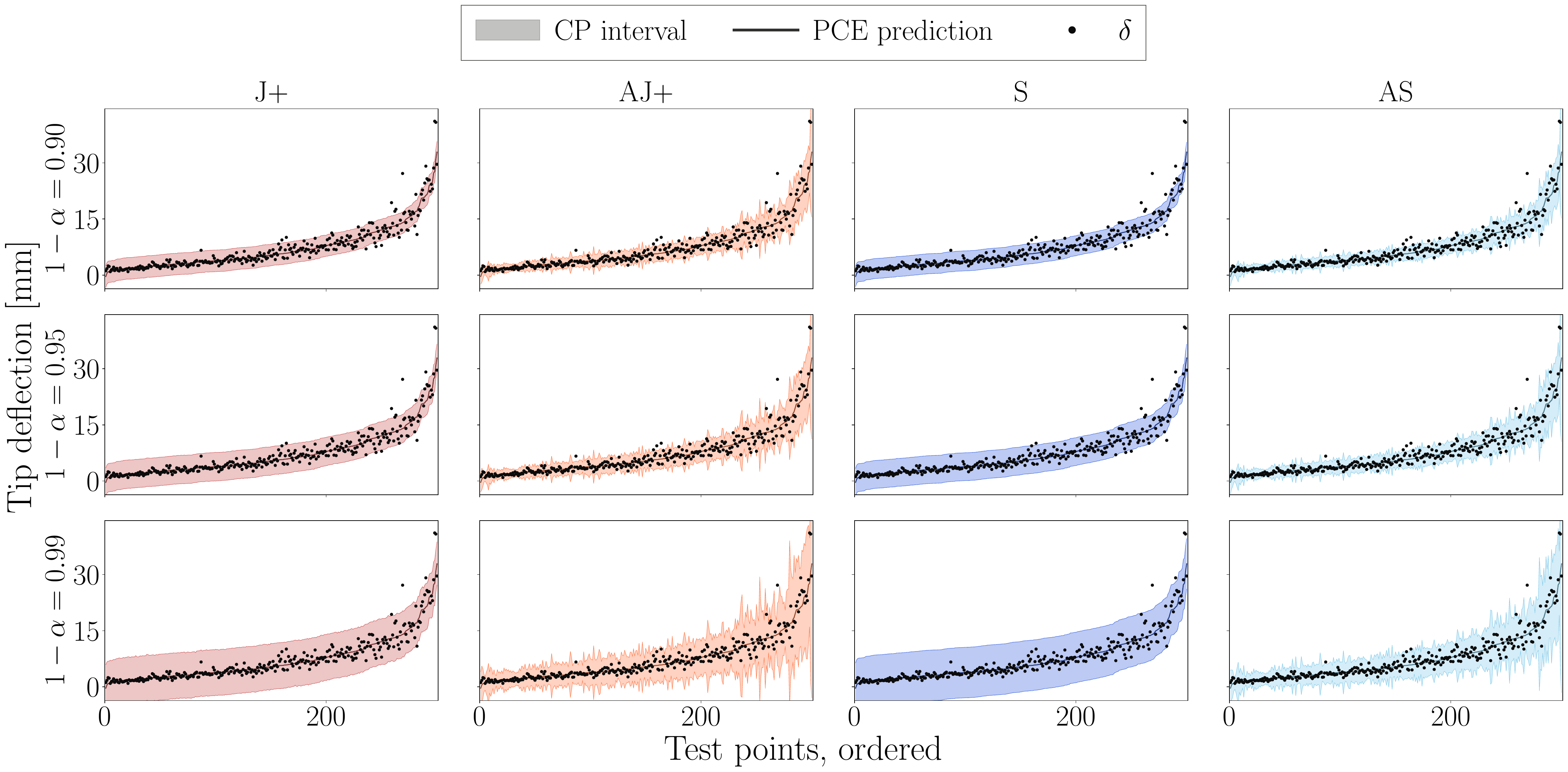}
\caption{Single-seed \gls{cp} intervals of all total-degree \gls{ols}-fitted \gls{cp}-\gls{pce} variants applied to the cantilever beam model. The test points shown (300 out of $10^3$) are ordered according to the predicted tip deflection value.}
\label{fig:cantilever-ordered-grid}
\end{figure}

The impact of adaptivity is further illustrated in Figure~\ref{fig:cantilever-ordered-grid}, showing \gls{cp} intervals for a representative seed out of the $N_{\mathrm{rep}}=100$ repetitions.
As can be observed, the non-adaptive variants yield nearly constant intervals irrespective of local prediction uncertainty, leading to severe under-coverage of high-deflection data (high noise), as well as to overly conservative intervals for low deflections (low noise). 
In contrast, the width of the adaptive intervals follows the local noise level, i.e., the intervals are narrow for small deflections and widen as the deflection increases.

\subsubsection{Sparse CP-PCE}
\label{sec:cantilever-sparse-pce}
We now consider a much smaller training dataset of size $M=150$ and employ sparse \glspl{pce} for both the base and the auxiliary models, computed with the hybrid \gls{lar} method~\cite{blatman2011adaptive}.
The polynomial degrees ($P_{\mathrm{base}}=3$, $P_{\mathrm{aux}}=2$) and $q$-norm values ($q_{\mathrm{base}}=q_{\mathrm{aux}}=1$) are selected using a grid-search approach.
The AJ+ variant is computed using the heuristic simplification detailed in Section~\ref{sec:acp-pce-sparse}, thus forgoing any coverage guarantee, whereas J+ is computed by explicitly training $M$ base models, thus retaining its $(1-2\alpha)$ marginal coverage guarantee.

Evaluated on the test dataset, the base model yields a coefficient of determination $R_{\text{test}}^2 = 0.90$, on average. 
The computation time of each \gls{cp}-\gls{pce} variant is: J+ in $2.62$~s, AJ+ in $3.47$~s, S in $0.26$~s, and AS in $0.53$~s.
The computation times of J+ and AJ+ are attributed predominantly to the explicit training of the $M$ \gls{loo} base models needed by both methods, whereas the $M$ sparse fits of the auxiliary \gls{pce} in AJ+ add only little computational overhead due to the small maximum polynomial degree.
Note that the guarantee-preserving approach of Section~\ref{sec:acp-pce-eta} would require $M(M+1)/2 = 11{,}325$ sparse fits of the base \gls{pce} model, increasing the computation time of AJ+ to approximately $3.3$~minutes, i.e., $\approx 57\times$ the cost of the heuristic, simplified variant.

\begin{figure}[t!]
    \centering
    \begin{subfigure}{0.7\textwidth}
        \centering
        \includegraphics[width=\linewidth]{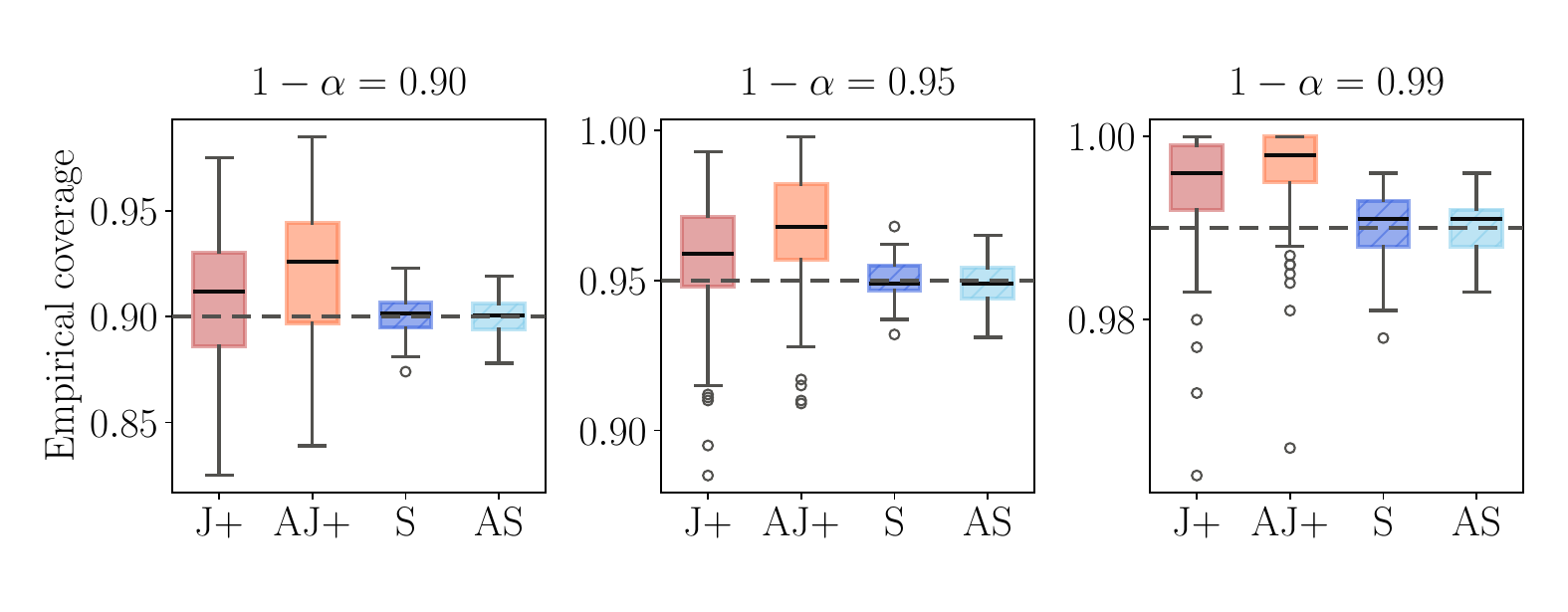}
        \caption{Empirical coverage.}
        \label{fig:cantilever-coverage-sparse}
    \end{subfigure}
    \begin{subfigure}{0.7\textwidth}
        \centering
        \includegraphics[width=\linewidth]{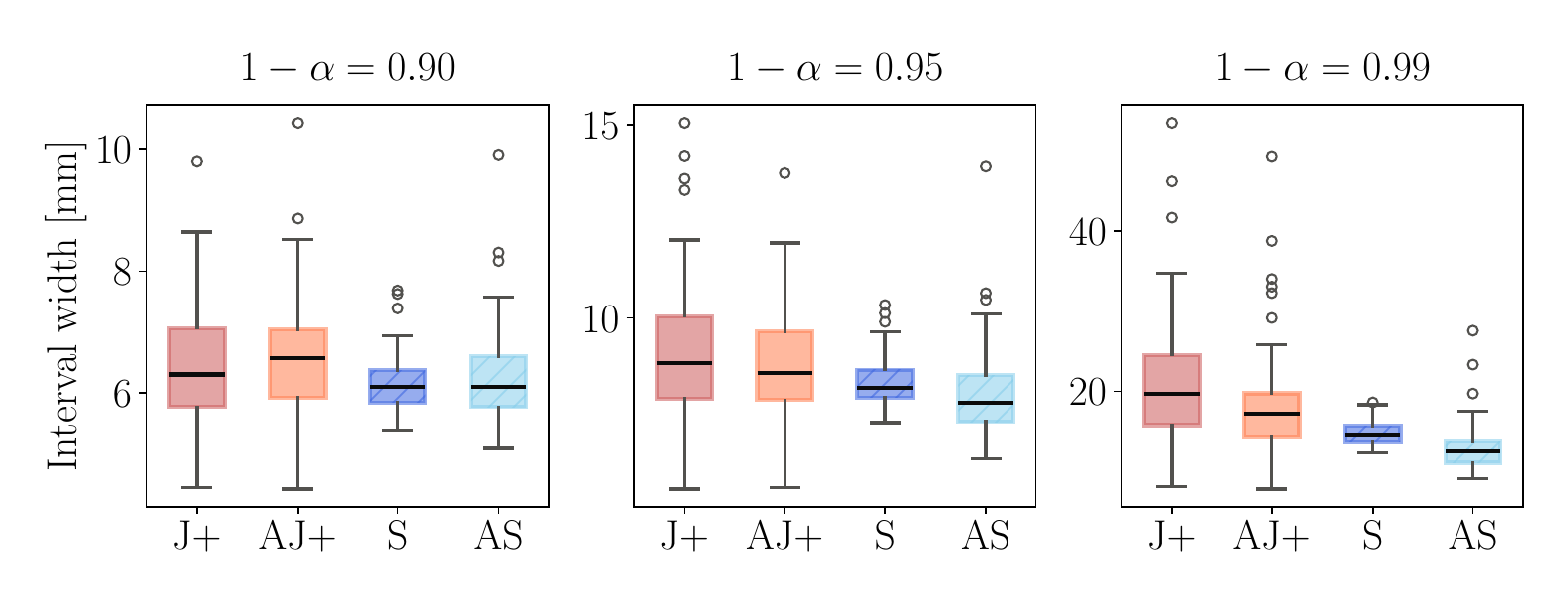}
        \caption{Mean interval width.}
        \label{fig:cantilever-intervals-sparse}
    \end{subfigure}
    \begin{subfigure}{0.7\textwidth}
        \centering
        \includegraphics[width=\linewidth]{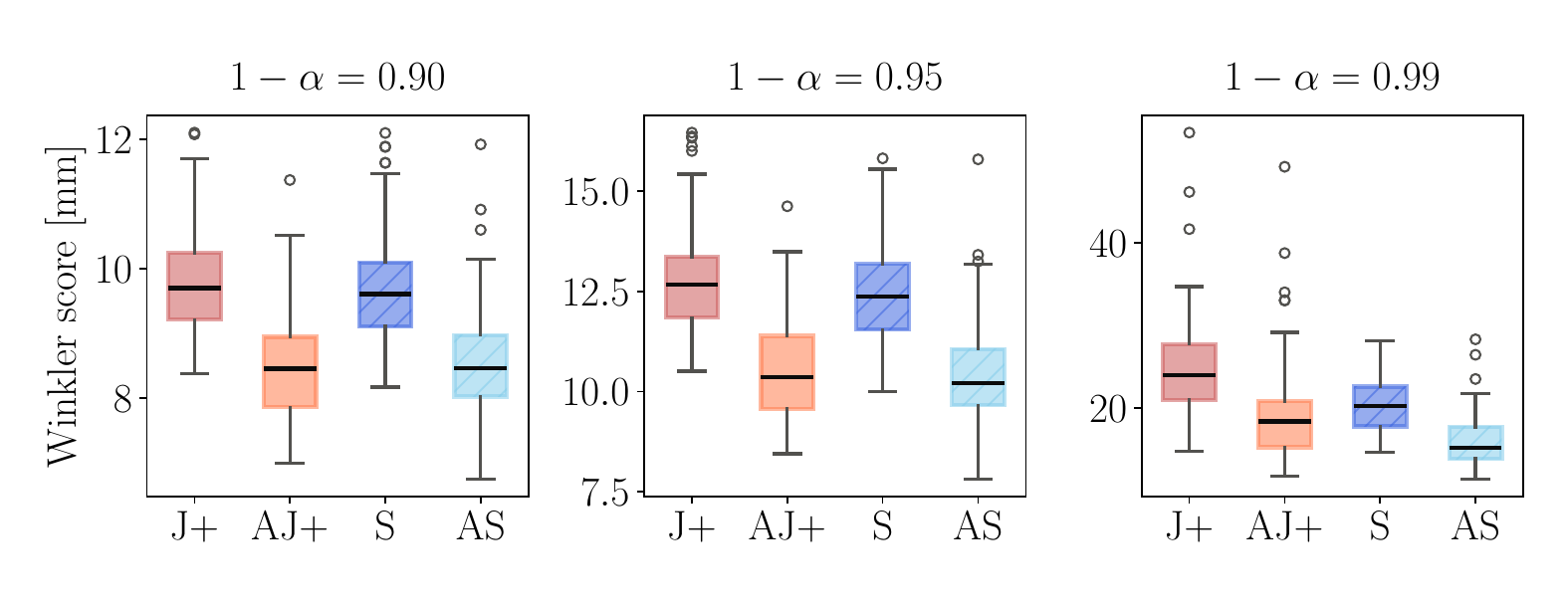}
        \caption{Winkler score.}
        \label{fig:cantilever-winkler-sparse}
    \end{subfigure}
    \caption{Box-plots of empirical coverage, mean interval width, and Winkler score by the different sparse \gls{cp}-\gls{pce} variants applied to the cantilever beam model.}
    \label{fig:cantilever-box-plots-sparse}
\end{figure}

\begin{figure}[t!]
\centering
\includegraphics[width=0.7\linewidth]{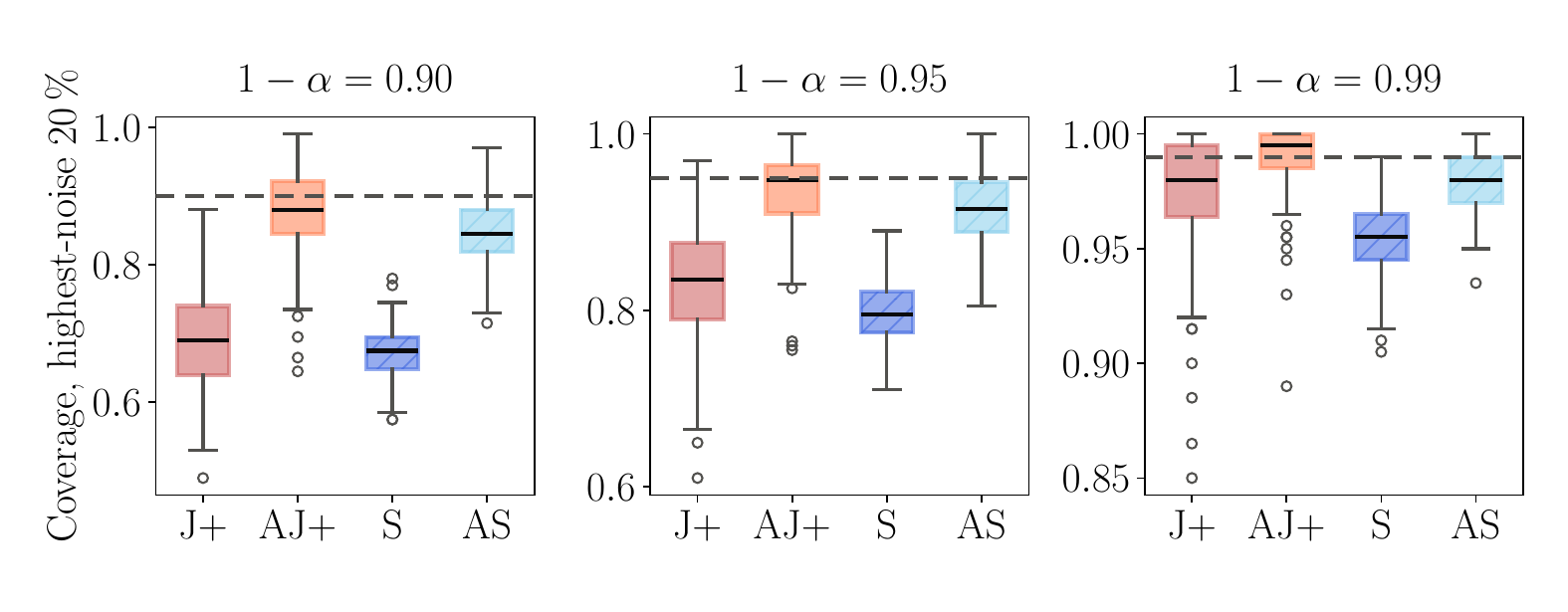}
\caption{Box-plots of the empirical coverage of the different sparse \gls{cp}-\gls{pce} variants on the 20\% of test points with the highest noise level, for the cantilever beam benchmark.}
\label{fig:cantilever-cov-high-sparse}
\end{figure}

Similar results as in Section~\ref{sec:cantilever-td-pce} are reported, where Figure~\ref{fig:cantilever-box-plots-sparse} presents the box-plot results for all evaluation metrics and Figure~\ref{fig:cantilever-cov-high-sparse} presents the highest-noise coverage results.
The single-seed results are omitted here.
Once again, all variants attain the target coverage, on average.
The coverages of S and AS are practically exact due to the $J=10^5$ calibration data, whereas J+ and AJ+ are slightly conservative and exhibit larger variations, since they are calibrated on $M=150$ data points only.
As in the total-degree case, AJ+ and AS yield clearly lower Winkler scores than J+ and S, namely by approximately 13\%, 17\%, and 23\% (AJ+ vs.\ J+) and 12\%, 16\%, and 22\% (AS vs.\ S) for $\alpha = 0.1, 0.05, 0.01$, respectively.
This improvement is obtained either without or with only a small reduction in the mean interval width, demonstrating that it stems from the redistribution of the interval width according to the noise level.
This is further verified by the empirical coverages on the 20\% of test points with the highest noise level, where J+ and S drop well below the target, whereas AJ+ and AS perform substantially better, especially the former.

\subsection{Engineering use case: Steam generator clogging dataset}
\label{sec:thyc}
This engineering use case concerns the problem of clogging in steam generators of pressurized water nuclear reactors~\cite{jaber2025sensitivity, jaber2025conformal}.
After prolonged operation, iron oxide deposits accumulate in the flow holes of the steam generator's tube support plates, thus reducing the available flow area.
The extent of this clogging is quantified by the clogging rate $\tau_{\text{c}}$, which is an average measure of the obstruction of the flow holes.

We employ a publicly available dataset of $10^3$ Monte Carlo simulations\footnote{\url{https://github.com/vincentblot28/conformalized_gp/tree/main/data/thyc}}, computed with the ``THYC-Puffer-DEPOTHYC'' simulation chain~\cite{jaber2025sensitivity}.
Each simulation provides the clogging rate $\tau_{\text{c}}$ at a fixed operating time for a random realization of $N = 7$ uncertain model parameters, which are modeled as mutually independent random variables with the distributions listed in Table~\ref{tab:thyc-inputs}.

\begin{table}[b!]
\centering
\caption{Input parameters of the THYC-Puffer-DEPOTHYC simulation chain and their probability distributions~\cite{jaber2025sensitivity}. $\mathcal{N}(m, s)$ denotes a Gaussian distribution with mean $m$ and standard deviation $s$. $\mathcal{T}(a, b, c)$ denotes a triangular distribution on $[a, c]$ with mode $b$.}
\label{tab:thyc-inputs}
\begin{tabular}{lll}
\hline
Parameter & Description & Distribution \\
\hline
$\kappa_1$ & First parameter of the clogging rate correlation & $\mathcal{N}(101.6, 4.0)$ \\
$\kappa_2$ & Second parameter of the clogging rate correlation & $\mathcal{N}(0.0233, 0.0005)$ \\
$\epsilon_{\text{e}}$ & Porosity of the fouling deposits & $\mathcal{T}(0.2, 0.3, 0.5)$ \\
$\epsilon_{\text{c}}$ & Porosity of the clogging deposits & $\mathcal{T}(0.01, 0.05, 0.3)$ \\
$d_{\text{p}}$ & Iron oxide particle diameter (m) & $\mathcal{T}(0.5, 5.0, 10.0) \times 10^{-6}$ \\
$\Gamma_{\text{p}}(0)$ & Initial mass fraction of solid particles & $\mathcal{T}(1.0, 4.5, 8.0) \times 10^{-9}$ \\
$a_{\text{v}}$ & Calibration parameter of the vena contracta mechanism & $\mathcal{T}(0.1, 7.8, 12.0) \times 10^{-4}$ \\
\hline
\end{tabular}
\end{table}

In this study, the available dataset is fixed and no new data can be generated. 
To assess the variability of the evaluation metrics, the numerical experiments are repeated $N_\mathrm{rep} = 100$ times by randomly reshuffling the original dataset before partitioning it into training and test data.
The size $M$ of the training dataset varies depending on the form and computation of the \gls{pce} models, i.e., total-degree \gls{ols}-fitted or sparse, whereas the test dataset has size $L=200$ throughout.
For the split \gls{cp}-\gls{pce} variants, the training dataset is further split into a \emph{proper} training dataset of size $M_{\mathrm{S}}$ and a calibration dataset of size $J$. 
Therefore, the split and jackknife+ variants do not share the same base \gls{pce} model in this case.
As before, three significance levels are tested, i.e., $\alpha \in \left\{0.1, 0.05, 0.01\right\}$.

Additionally, to quantify adaptivity in the absence of a known noise level, we report Spearman's rank correlation coefficient $r_{\text{S}} \in \left[-1,1,\right]$ between the interval widths $\left(u^{(l)} - \ell^{(l)}\right)_{l=1}^L$ and the absolute prediction errors $\left(|y^{(l)} - \widehat\mu(\mathbf{x}^{(l)})|\right)_{l=1}^L$ on the test dataset. 
Positive values indicate that wider intervals are assigned where the errors are larger. 
Note that for non-adaptive split \gls{cp}, the width is constant and $r_{\text{S}}$ is undefined.

\subsubsection{Total-degree OLS-fitted CP-PCE}
\label{sec:thyc-td-pce}
In this case, both the base model and the auxiliary model are total-degree \gls{ols}-fitted \glspl{pce}, where the best results were obtained with $P_\text{base}=3$ and $P_\text{aux}=2$.
The adaptive jackknife+ \gls{cp} intervals are therefore computed with the guarantee-preserving method of Section~\ref{sec:acp-pce-eta}, using the closed-form expressions of Section~\ref{sec:acp-pce-fixed-ols}.
The original dataset is randomly reshuffled and partitioned into $M = 800$ training points and $L=200$ test points.
For S and AS, the original training dataset is further partitioned into a proper training dataset of size $M_\mathrm{S}=500$ and a calibration dataset of size $J=300$.
Evaluated on the test dataset, all base \gls{pce} models yield a coefficient of determination $R_{\text{test}}^2 \approx 0.98$ despite the difference in the training data.
Owing to the closed-form expressions, all \gls{cp}-\gls{pce} variants are computed very fast: J+ in $0.027$~s, AJ+ in $0.19$~s, S in $0.014$~s, and AS in $0.074$~s, on average.

\begin{figure}[t!]
    \centering
    \begin{subfigure}{0.7\textwidth}
        \centering
        \includegraphics[width=\linewidth]{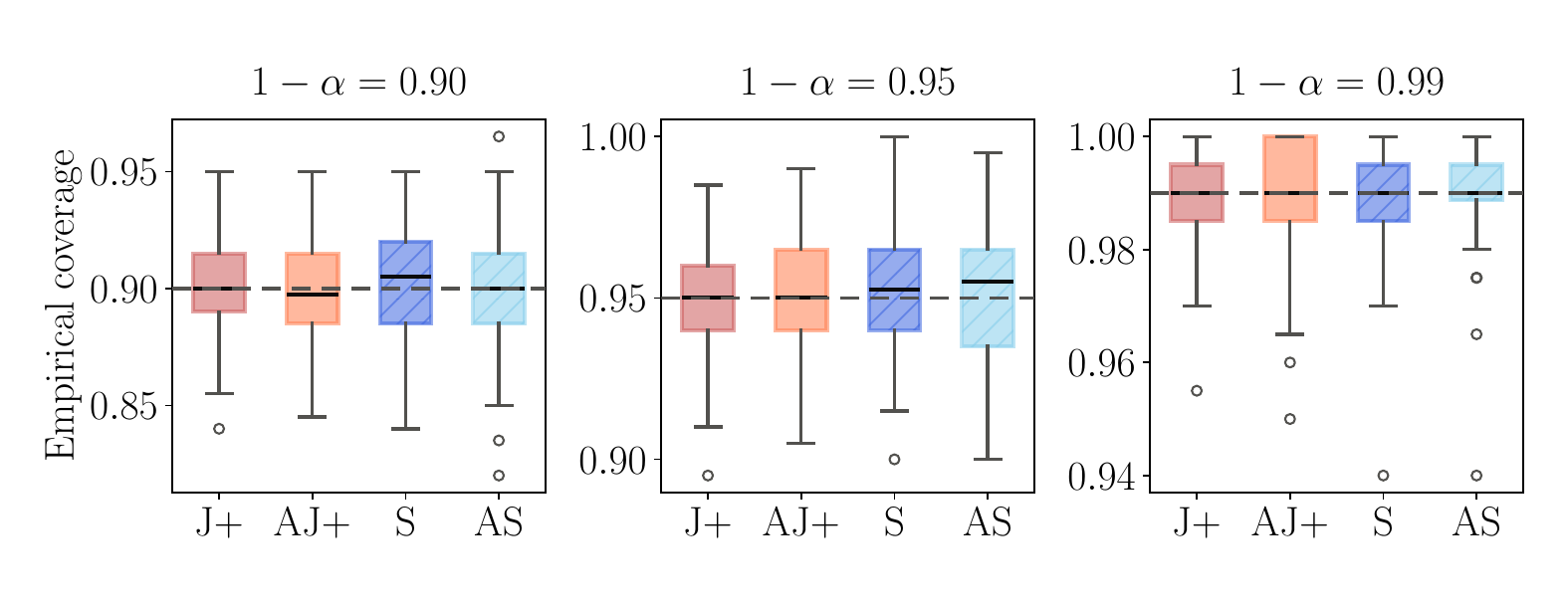}
        \caption{Empirical coverage.}
        \label{fig:thyc-coverage}
    \end{subfigure}
    \begin{subfigure}{0.7\textwidth}
        \centering
        \includegraphics[width=\linewidth]{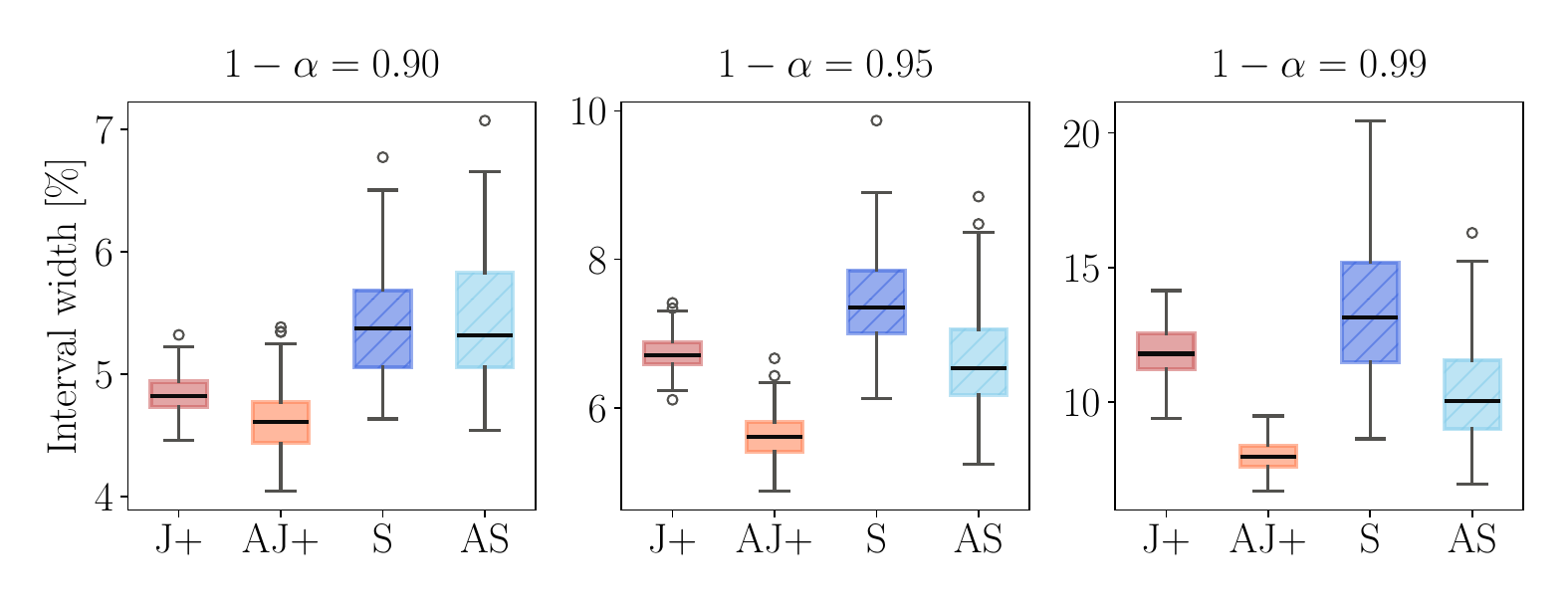}
        \caption{Mean interval width.}
        \label{fig:thyc-intervals}
    \end{subfigure}
    \begin{subfigure}{0.7\textwidth}
        \centering
        \includegraphics[width=\linewidth]{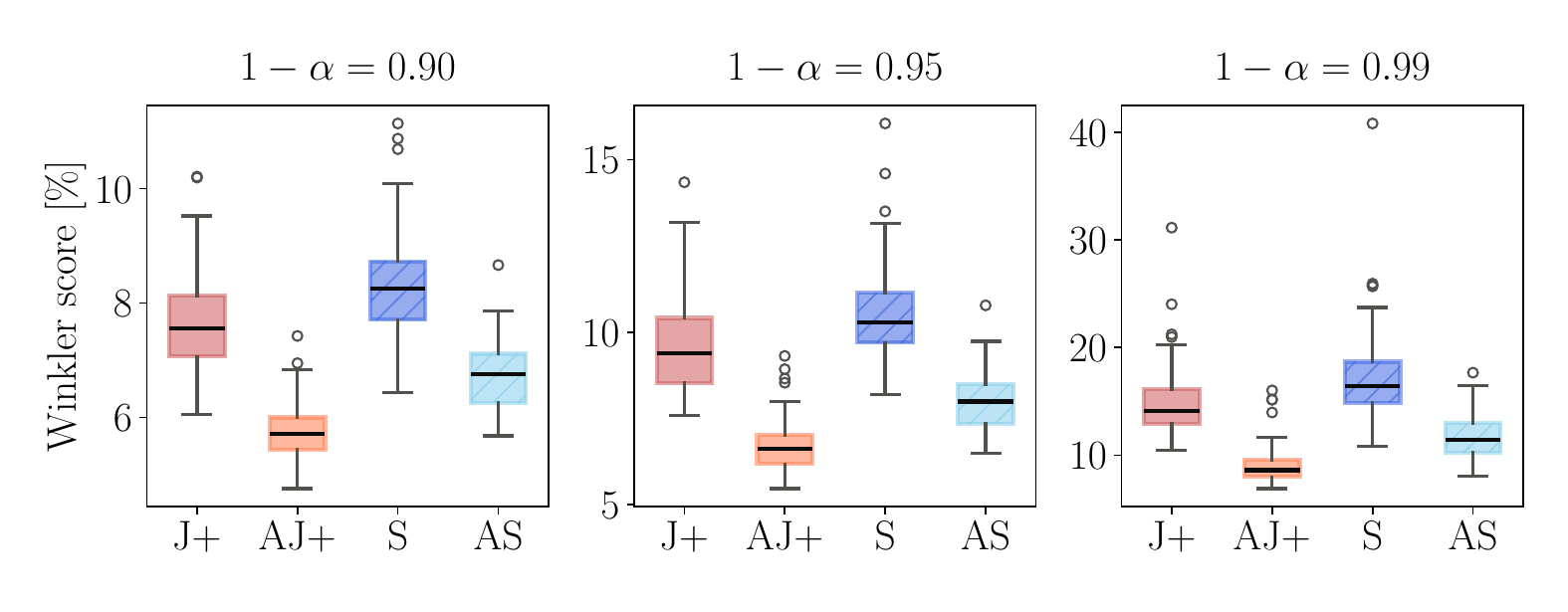}
        \caption{Winkler score.}
        \label{fig:thyc-winkler}
    \end{subfigure}
    \caption{Box-plots of empirical coverage, mean interval width, and Winkler score by the different total-degree \gls{ols}-fitted \gls{cp}-\gls{pce} variants applied to the steam generator clogging use case.}
    \label{fig:thyc-box-plots}
\end{figure}

Figure~\ref{fig:thyc-box-plots} shows the results for all metrics, target coverages, and \gls{cp}-\gls{pce} variants, across the $N_{\mathrm{rep}}=100$ repetitions.
While all variants attain the target coverage, they differ considerably in statistical efficiency.
Local adaptivity improves both the jackknife+ and the split \gls{cp}-\glspl{pce}, yielding narrower intervals for the same coverage. 
Comparing J+ and AJ+, the latter reduces the mean interval width by $4\%$, $17\%$, and $32\%$, and the Winkler score by $24\%$, $30\%$, and $39\%$ for $\alpha = 0.1, 0.05, 0.01$, respectively.
Adaptivity yields similar gains for split \gls{cp} as well.
Overall, AJ+ is the most efficient variant.
Its advantage over AS is attributed mainly to the smaller training dataset of the latter and to its calibration on only $J=300$ points.
This underlines the advantage of jackknife+ \gls{cp} methods under limited data availability.

\begin{figure}[t!]
\centering
 \includegraphics[width=\linewidth]{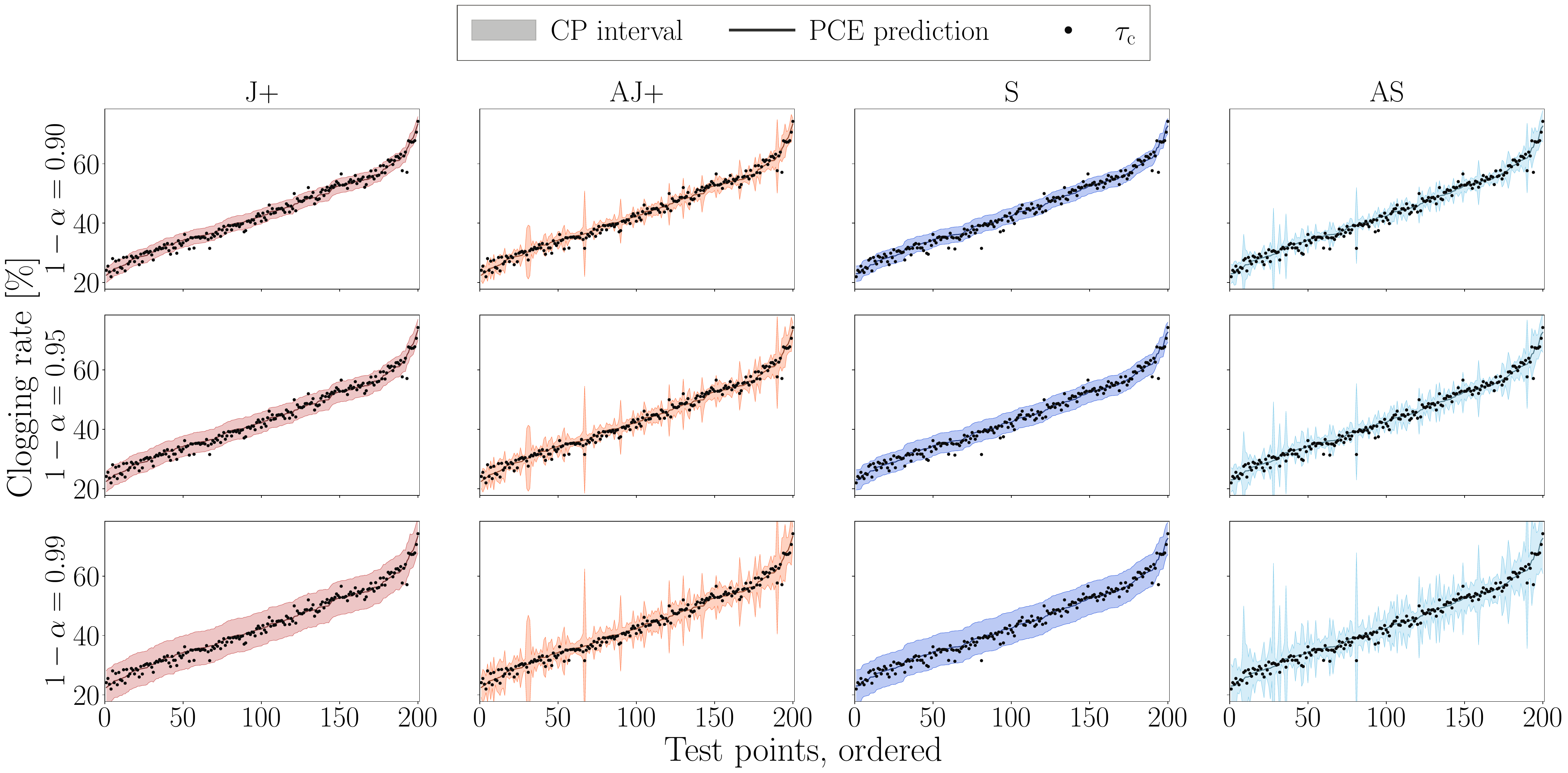}    
 \caption{Single-seed \Gls{cp} intervals of all total-degree \gls{ols}-fitted \gls{cp}-\gls{pce} variants applied to the steam generator clogging use case. The $L=200$ test points are ordered according to the predicted clogging rate value.}
 \label{fig:thyc-parity-plots}
\end{figure}

The impact of local adaptivity is further illustrated in Figure~\ref{fig:thyc-parity-plots}, showing \gls{cp} intervals for a representative seed out of the $N_{\mathrm{rep}}=100$ repetitions. 
As can be observed, the non-adaptive variants provide intervals of nearly constant width for all predictions, whereas the widths of the adaptive intervals vary with the local prediction uncertainty. 
Moreover, the adaptive intervals widen mainly where the non-adaptive ones fail to provide coverage, for example comparing the two jackknife-based \gls{cp}-\glspl{pce} for $1-\alpha=0.90$.
The split \gls{cp}-\gls{pce} variants behave similarly.
Although corresponding to a single seed, these results are in general agreement with the aggregated results shown in Figure~\ref{fig:thyc-intervals}. 

\begin{figure}[t!]
\centering
\includegraphics[width=0.7\linewidth]{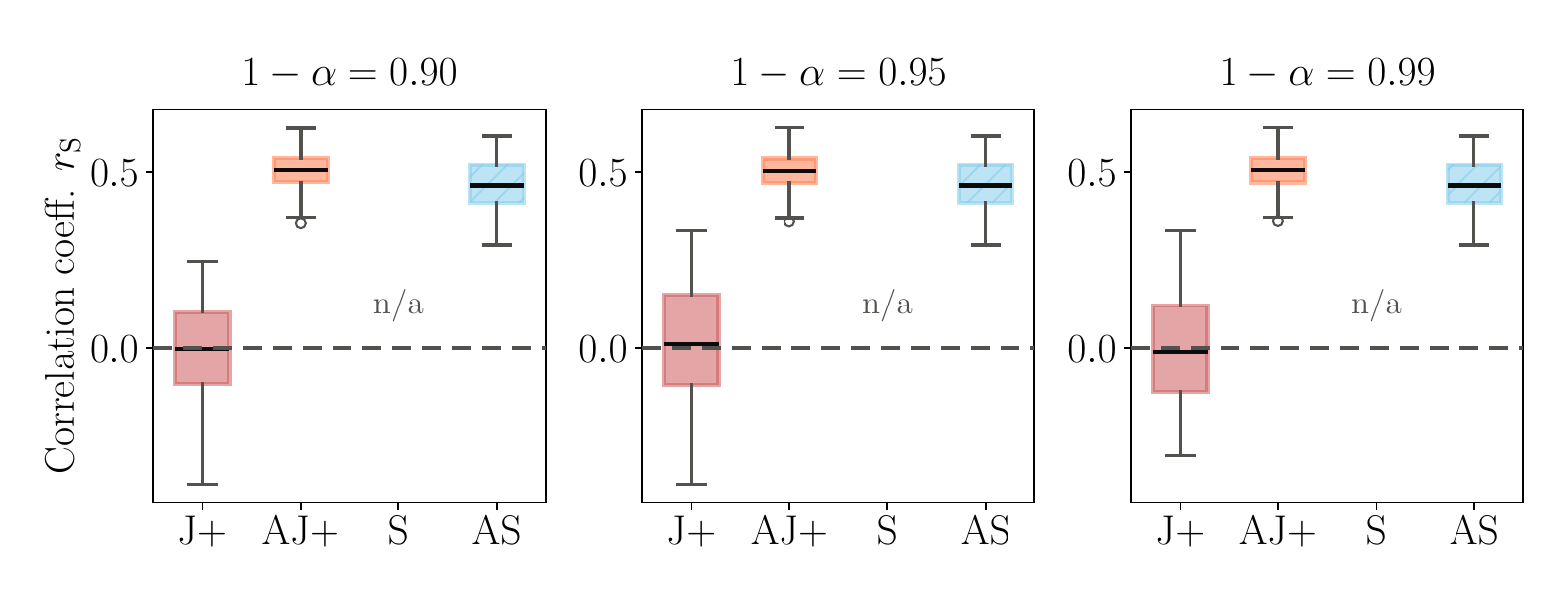}
\caption{Box-plots of Spearman's rank correlation coefficient by the different total-degree \gls{ols}-fitted \gls{cp}-\gls{pce} variants applied to the team generator clogging use case.}
\label{fig:thyc-spearman}
\end{figure}

The adaptivity of the intervals is also reflected in the Spearman rank
correlation coefficient results presented in Figure~\ref{fig:thyc-spearman}.
A clear positive correlation is evident for both adaptive \gls{cp}-\gls{pce} variants, with medians of $r_{\text{S}}^{\text{AJ+}} \approx 0.50$ and $r_{\text{S}}^{\text{AS}} \approx 0.46$ at all target coverage levels.
In contrast, J+ shows no systematic correlation, with $r_{\text{S}}^{\text{J+}}$ scattering around zero. 

\subsubsection{Sparse CP-PCE}
\label{sec:thyc-sparse-pce}
We now consider a much smaller training dataset of size $M=200$.
For each of the $N_{\mathrm{rep}}= 100$ repetitions, the $L=200$ test points are the same as in Section~\ref{sec:thyc-td-pce}, and the training points are the first $M=200$ of the corresponding $800$ previous training points, so that the two studies are paired.
Both the base and auxiliary models are now sparse \glspl{pce} computed with the hybrid \gls{lar} method, where the maximum polynomial degrees ($P_{\mathrm{base}}=6$, $P_{\mathrm{aux}}=2$) and $q$-norms ($q_{\mathrm{base}}=0.6$, $q_{\mathrm{aux}}=0.9$) were chosen using a grid-search approach.
The closed-form expressions from Section~\ref{sec:acp-pce-fixed-ols} are not applicable in this setting.
The J+ \gls{cp}-\gls{pce} is computed by explicitly retraining $M$ models, thus retaining its $(1-2\alpha)$ coverage guarantee.
The AJ+ \gls{cp}-\gls{pce} is computed with the heuristic simplification introduced in Section~\ref{sec:acp-pce-sparse}, thus forgoing the coverage guarantee.
Due to the limited data availability, the split \gls{cp} methods are not employed in this case.

On average, the base \gls{pce} model yields a coefficient of determination $R_{\text{test}}^2=0.97$ on the test dataset, and the standard and adaptive jackknife+ \gls{cp}-\glspl{pce} are computed in $10.96$~s and $12.21$~s, respectively.
This cost is dominated by the $M$ explicitly trained \gls{loo} base models, which are needed by both variants. 
The guarantee-preserving approach would instead require $M(M+1)/2 = 20{,}100$ explicitly trained base \gls{pce} models, resulting in a training time of $\approx19$ minutes, i.e., $\approx 90\times$ the cost of the heuristic, simplified variant.

\begin{figure}[t!]
    \centering
    \begin{subfigure}{0.7\textwidth}
        \centering
        \includegraphics[width=1\linewidth]{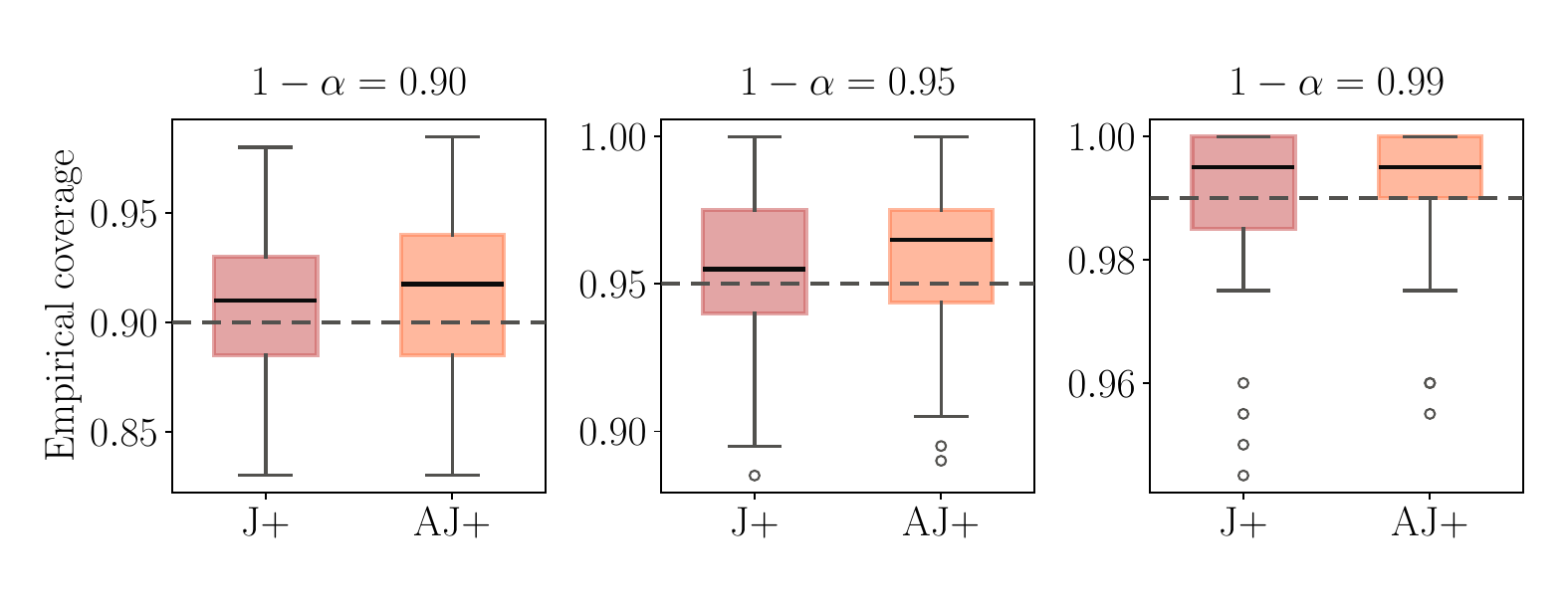}
        \caption{Empirical coverage.}
        \label{fig:thyc-coverage-sparse}
    \end{subfigure}
    \begin{subfigure}{0.7\textwidth}
        \centering
        \includegraphics[width=1\linewidth]{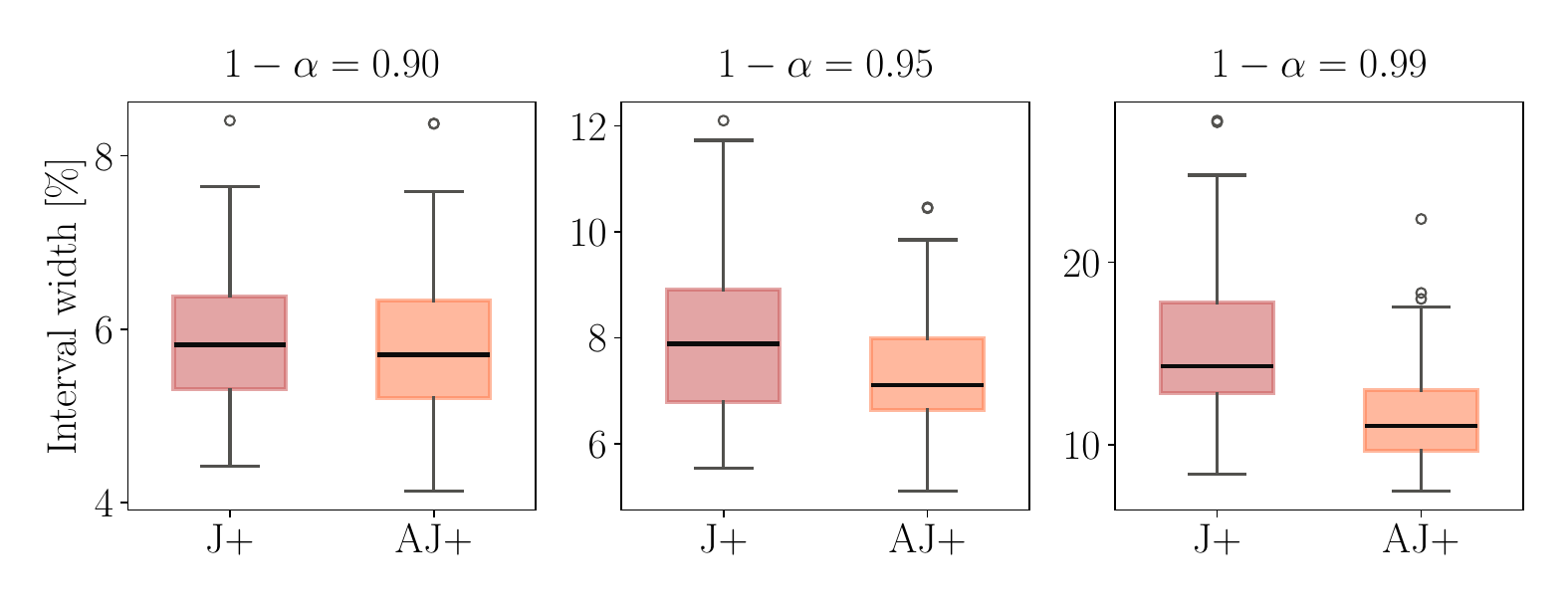}
        \caption{Mean interval width.}
        \label{fig:thyc-intervals-sparse}
    \end{subfigure}
    \begin{subfigure}{0.7\textwidth}
        \centering
        \includegraphics[width=1\linewidth]{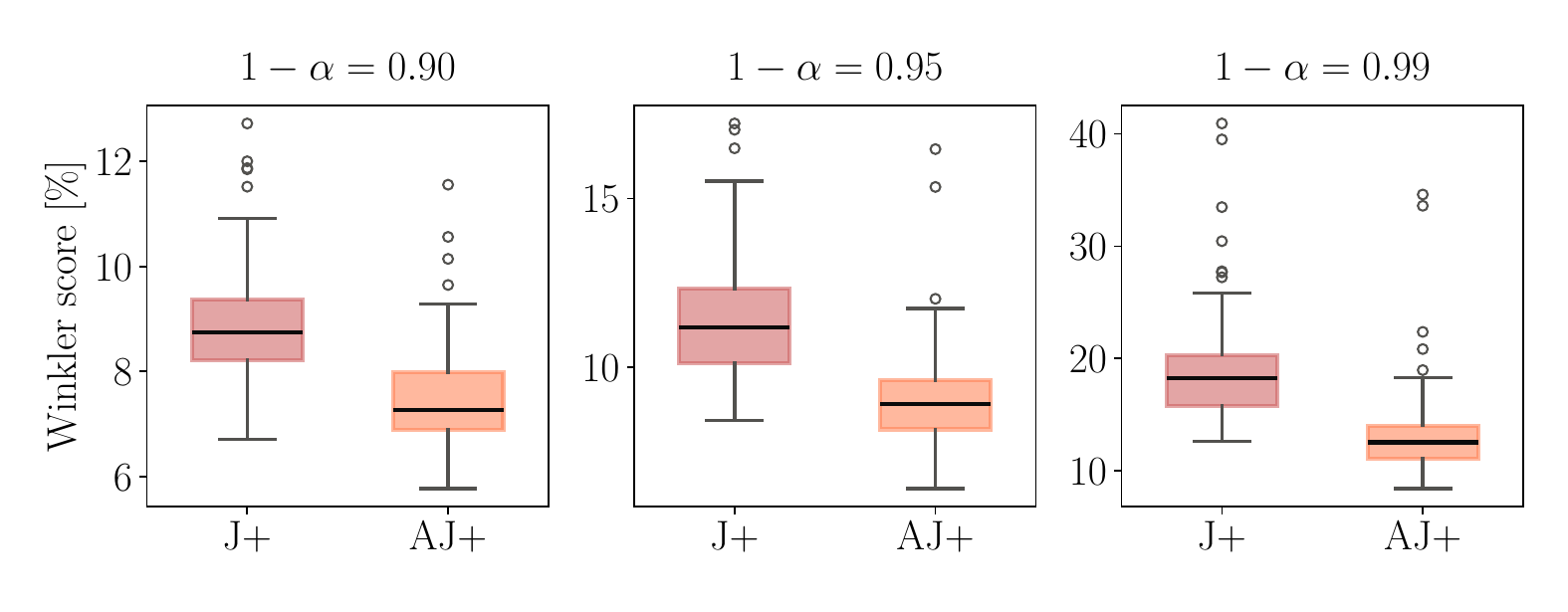}
        \caption{Winkler score.}
        \label{fig:thyc-winkler-sparse}
    \end{subfigure}
    \caption{Box-plots of empirical coverage, mean interval width, and Winkler score by jackknife+ sparse \gls{cp}-\glspl{pce} applied to the steam generator clogging use case.}
    \label{fig:thyc-box-plots-sparse}
\end{figure}

\begin{figure}[t!]
\centering
 \includegraphics[width=0.7\textwidth]{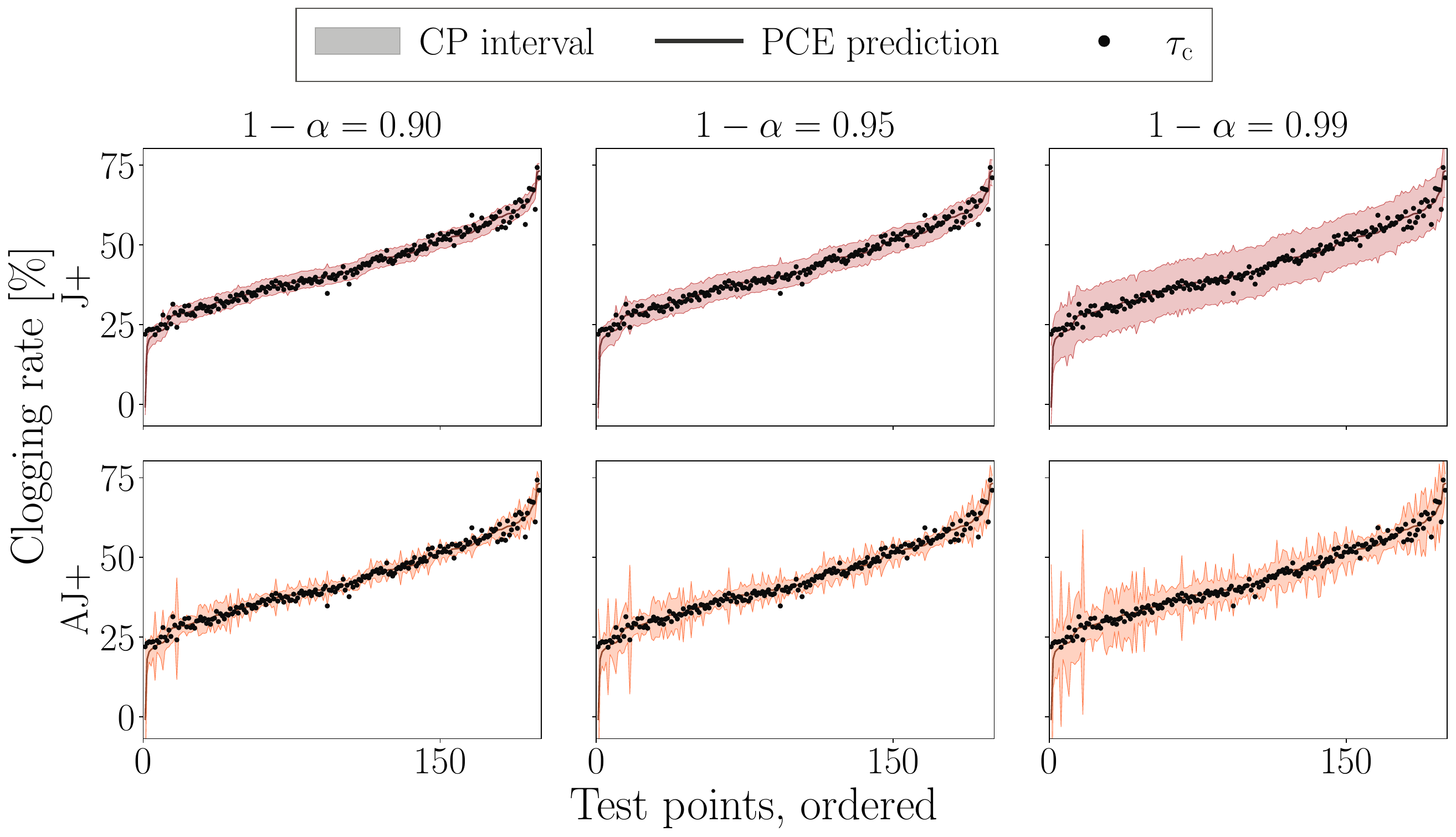}    
 \caption{Single-seed \Gls{cp} intervals of jackknife+ sparse \gls{cp}-\glspl{pce} applied to the steam generator clogging use case. The $L=200$ test points are ordered according to the predicted clogging rate value.}
 \label{fig:thyc-parity-plots-sparse}
\end{figure}

\begin{figure}[t!]
\centering
\includegraphics[width=0.7\linewidth]{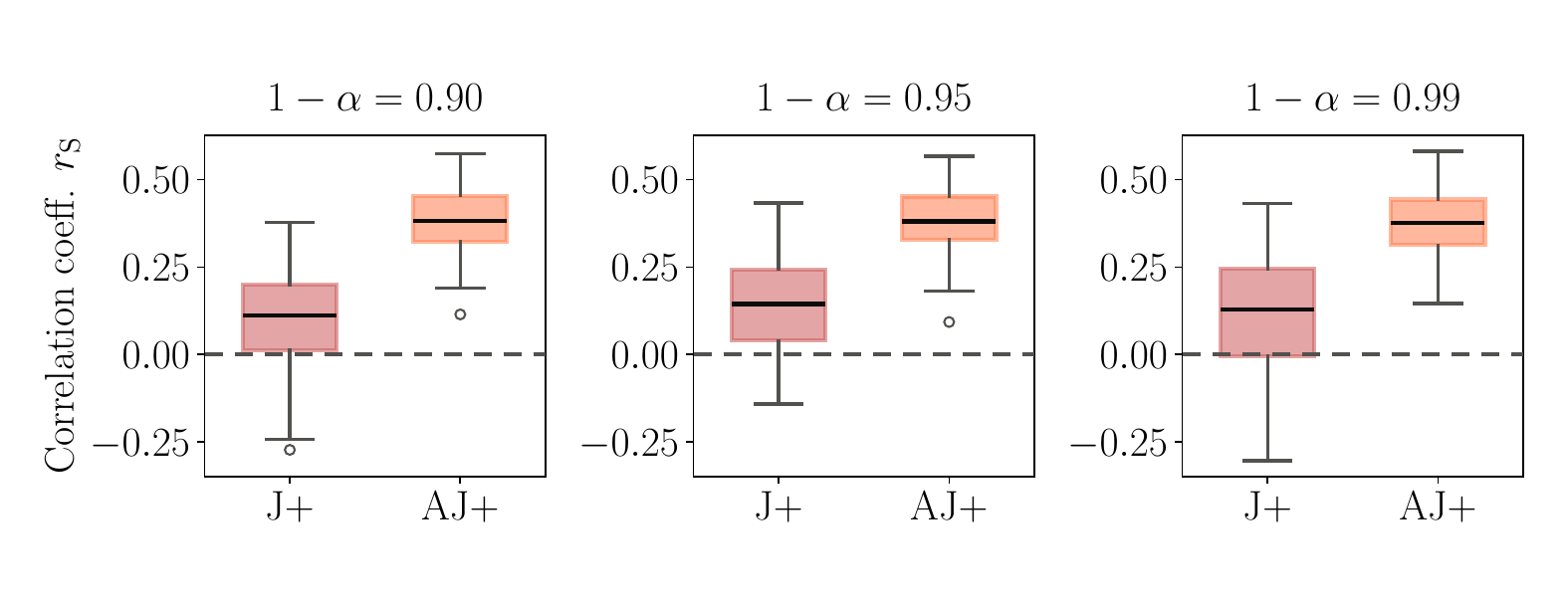}
\caption{Box-plots of Spearman's rank correlation coefficient by jackknife+ sparse \gls{cp}-\gls{pce} variants applied to the steam generator clogging use case.}
\label{fig:thyc-spearman-sparse}
\end{figure}

Figure~\ref{fig:thyc-box-plots-sparse} presents the coverage, interval width, and Winkler score results over the $N_{\mathrm{rep}}=100$ repetitions. 
Both J+ and AJ+ \gls{cp}-\glspl{pce} reach the prescribed target coverages. 
However, the adaptive variant is clearly more efficient, yielding both tighter intervals and lower Winkler scores.
These advantages can also be observed in the single-seed \gls{cp} interval results shown in Figure~\ref{fig:thyc-parity-plots-sparse}, where it is evident that the nearly constant non-adaptive intervals are typically too conservative compared to the adaptive ones.
Interval adaptivity is further reflected in the Spearman rank
correlation coefficient results presented in Figure~\ref{fig:thyc-spearman-sparse}. 
A clear positive correlation exists for AJ+, with a median of
$r_{\text{S}}^{\text{AJ+}} \approx 0.38$ at all target coverage levels. 
For J+, a weak positive correlation can be observed (median $r_{\text{S}}^{\text{J+}} \approx 0.1$), however, with a large spread across the repetitions that includes negative values, essentially indicating little to no correlation in practice.

\section{Summary and conclusions}
\label{sec:conclusion}
This work introduced an adaptive jackknife+ \gls{cp} framework for \gls{pce} regression, which complements \gls{pce} predictions with prediction intervals whose widths adapt to the local, input-dependent prediction uncertainty.
The adaptivity is achieved through normalized non-conformity scores, where the local uncertainty estimator is an auxiliary \gls{pce} trained on the log-transformed \gls{loo} residuals of the base \gls{pce}.
To retain the $(1-2\alpha)$ finite-sample marginal coverage guarantee of jackknife+ \gls{cp}, the \gls{loo} auxiliary models must be trained on nested \gls{loo} residuals, obtained from models that omit two training points at a time.
For fixed-basis, \gls{ols}-fitted \glspl{pce}, closed-form expressions can be used for all required quantities.
The guarantee-preserving adaptive intervals are thus obtained from a single \gls{ols} fit of the base model, without any explicit retraining.
For sparse \glspl{pce}, where the basis selection depends on the training data, these closed-form expressions are not applicable and the guarantee-preserving approach requires $M(M+1)/2$ sparse fits of the base model.
To mitigate the computational cost, we proposed a heuristic simplification that reduces the overhead of adaptivity to $M$ sparse fits of the auxiliary model, at the price of forgoing the formal coverage guarantee.

The numerical experiments on a cantilever beam benchmark with known heteroscedastic noise and on an engineering dataset of steam generator clogging simulations confirmed the benefits of the suggested approach.
All adaptive and non-adaptive \gls{cp}-\gls{pce} variants attained the target coverage on average, including the adaptive jackknife+ variant for sparse \glspl{pce}.
Compared to its non-adaptive counterpart, the adaptive jackknife+ \gls{cp}-\gls{pce} reduced the Winkler score substantially, indicating that the interval width was successfully redistributed according to the local prediction uncertainty. 
These gains are evident also in cases where the mean interval width is not reduced or reduced only marginally.
In particular, the non-adaptive intervals severely under-covered the test points with the highest prediction uncertainty, whereas the adaptive intervals retained coverages close to the target level for these points as well.
In the benchmark use case, the adaptive jackknife+ \gls{cp}-\gls{pce} performed on par with the adaptive split \gls{cp}-\gls{pce}, where a very large calibration dataset was used for the latter.
In the engineering use case, where the available data must be shared between training and calibration for split \gls{cp}, the adaptive jackknife+ was clearly the better method, underlining the data efficiency of jackknife-based \gls{cp} in small-data regimes.

Some limitations of the suggested approach should be noted.
First, the provided coverage is marginal and adaptivity improves conditional coverage only to the extent that the auxiliary model captures the local prediction uncertainty.
This was evident for the sparse \glspl{pce} trained on small datasets, where the observed improvements were smaller.
Second, for small training datasets and high target coverages, the jackknife+ intervals are dominated by extreme non-conformity scores and thus become wide and variable, as observed for $M=150$ and $1-\alpha=0.99$ in the benchmark.
Third, the guarantee-preserving approach remains computationally prohibitive for sparse \glspl{pce}, hence, the coverage of the corresponding adaptive intervals is supported by empirical numerical evidence only.
Where possible, we aim to address these shortcomings in dedicated follow-up works.

\paragraph{Acknowledgments:} 
Arthur Hatstatt was supported by the Swiss National Science Foundation (SNSF) under grant No. 220074 (Project ORACLES: Optimization, Reliability And Calibration using Emulators of Stochastic computational models).
Dimitris Giovanis acknowledges support by the U.S. Department of Energy under Grant No.~DE-SC0024162.



\bibliographystyle{unsrt}
\bibliography{references}

@article{arbel2026primer,
  title={A primer on {Bayesian} neural networks: review and debates},
  author={Arbel, Julyan and Pitas, Konstantinos and Vladimirova, Mariia and Fortuin, Vincent},
  journal={Statistical Science},
  volume={41},
  number={2},
  pages={316--353},
  year={2026},
  publisher={Institute of Mathematical Statistics}
}

@article{barber2021predictive,
  title={Predictive inference with the jackknife+},
  author={Barber, Rina Foygel and Candes, Emmanuel J and Ramdas, Aaditya and Tibshirani, Ryan J},
  journal={The Annals of Statistics},
  volume={49},
  number={1},
  pages={486--507},
  year={2021},
  publisher={JSTOR}
}

@article{barber2021limits,
  title={The limits of distribution-free conditional predictive inference},
  author={Barber, Rina Foygel and Candes, Emmanuel J and Ramdas, Aaditya and Tibshirani, Ryan J},
  journal={Information and Inference: A Journal of the IMA},
  volume={10},
  number={2},
  pages={455--482},
  year={2021},
  publisher={Oxford University Press}
}

@incollection{baudin2017openturns,
author="Baudin, Micha{\"e}l
and Dutfoy, Anne
and Iooss, Bertrand
and Popelin, Anne-Laure",
editor="Ghanem, Roger
and Higdon, David
and Owhadi, Houman",
title="{OpenTURNS}: An industrial software for uncertainty quantification in simulation",
bookTitle="Handbook of Uncertainty Quantification",
year="2017",
publisher="Springer International Publishing",
address="Cham",
pages="2001--2038",
}

@article{blatman2011adaptive,
  title={Adaptive sparse polynomial chaos expansion based on least angle regression},
  author={Blatman, G{\'e}raud and Sudret, Bruno},
  journal={Journal of Computational Physics},
  volume={230},
  number={6},
  pages={2345--2367},
  year={2011},
  publisher={Elsevier}
}

@phdthesis{blatman2009phd,
  author       = {Blatman, G{\'e}raud},
  title        = {Adaptive Sparse Polynomial Chaos Expansions for
                  Uncertainty Propagation and Sensitivity Analysis},
  school       = {Universit{\'e} Blaise Pascal, Clermont-Ferrand II},
  address      = {Clermont-Ferrand, France},
  year         = {2009},
  type         = {PhD thesis}
}

@article{chen2025machine,
  title={Machine learning as {CFD} surrogate models for rapid prediction of building-related physical fields: A review of methods and state-of-the-art},
  author={Chen, Tianle and Li, Ruibin and Hu, Xuesong and Zhang, Bingyang and Liu, Yuanyuan and Wang, Liangzhu Leon and Gao, Naiping},
  journal={Building and Environment},
  pages={113667},
  year={2025},
  publisher={Elsevier}
}

@article{deutschmann2024adaptive,
  title={Adaptive conformal regression with split-jackknife+ scores},
  author={Deutschmann, Nicolas and Rigotti, Mattia and Martinez, Maria Rodriguez},
  journal={Transactions on Machine Learning Research},
  year={2024}
}

@article{dheur2025unified,
  title={A unified comparative study with generalized conformity scores for multi-output conformal regression},
  author={Dheur, Victor and Fontana, Matteo and Estievenart, Yorick and Desobry, Naomi and Taieb, Souhaib Ben},
  journal={arXiv preprint arXiv:2501.10533},
  year={2025}
}

@inproceedings{gazin2024transductive,
  title={Transductive conformal inference with adaptive scores},
  author={Gazin, Ulysse and Blanchard, Gilles and Roquain, Etienne},
  booktitle={International Conference on Artificial Intelligence and Statistics},
  pages={1504--1512},
  year={2024},
  organization={PMLR}
}

@book{ghanem2003stochastic,
  title={Stochastic finite elements: a spectral approach},
  author={Ghanem, Roger G and Spanos, Pol D},
  year={2003},
  publisher={Courier Corporation}
}

@article{gneiting2007strictly,
  title={Strictly proper scoring rules, prediction, and estimation},
  author={Gneiting, Tilmann and Raftery, Adrian E},
  journal={Journal of the American statistical Association},
  volume={102},
  number={477},
  pages={359--378},
  year={2007},
  publisher={Taylor \& Francis}
}

@article{hatstatt2026conformal,
  title={Conformal prediction for full and sparse polynomial chaos expansions},
  author={Hatstatt, Arthur and Zhu, Xujia and Sudret, Bruno},
  journal={Computer Methods in Applied Mechanics and Engineering},
  volume={461},
  pages={119244},
  year={2026},
  publisher={Elsevier}
}

@article{jaber2025sensitivity,
  title={Sensitivity analyses of a multiphysics long-term clogging model for steam generators},
  author={Jaber, Edgar and Chabridon, Vincent and Remy, Emmanuel and Baudin, Micha{\"e}l and Lucor, Didier and Mougeot, Mathilde and Iooss, Bertrand},
  journal={International Journal for Uncertainty Quantification},
  volume={15},
  number={1},
  year={2025},
  publisher={Begel House Inc.}
}

@article{jaber2025conformal,
  title={Conformal approach to {Gaussian} process surrogate evaluation with marginal coverage guarantees},
  author={Jaber, Edgar and Blot, Vincent and Brunel, Nicolas JB and Chabridon, Vincent and Remy, Emmanuel and Iooss, Bertrand and Lucor, Didier and Mougeot, Mathilde and Leite, Alessandro},
  journal={Journal of Machine Learning for Modeling and Computing},
  volume={6},
  number={3},
  year={2025},
  publisher={Begel House Inc.}
}

@article{jakeman2015enhancing,
  title={Enhancing $\ell$1-minimization estimates of polynomial chaos expansions using basis selection},
  author={Jakeman, John D and Eldred, Michael S and Sargsyan, Khachik},
  journal={Journal of Computational Physics},
  volume={289},
  pages={18--34},
  year={2015},
  publisher={Elsevier}
}

@incollection{jiang2019surrogate,
  title={Surrogate-model-based design and optimization},
  author={Jiang, Ping and Zhou, Qi and Shao, Xinyu},
  booktitle={Surrogate Model-Based Engineering Design and Optimization},
  pages={135--236},
  year={2019},
  publisher={Springer}
}

@article{kato2023review,
  title={A review of nonconformity measures for conformal prediction in regression},
  author={Kato, Yuko and Tax, David MJ and Loog, Marco},
  journal={Conformal and probabilistic prediction with applications},
  pages={369--383},
  year={2023},
  publisher={PMLR}
}

@article{kersaudy2015new,
  title={A new surrogate modeling technique combining Kriging and polynomial chaos expansions--Application to uncertainty analysis in computational dosimetry},
  author={Kersaudy, Pierric and Sudret, Bruno and Varsier, Nad{\`e}ge and Picon, Odile and Wiart, Joe},
  journal={Journal of Computational Physics},
  volume={286},
  pages={103--117},
  year={2015},
  publisher={Elsevier}
}

@inproceedings{kharazian2024copal,
  title={{CoPAL}: Conformal Prediction for Active Learning with Application to Remaining Useful Life Estimation in Predictive Maintenance},
  author={Kharazian, Zahra and Lindgren, Tony and Magnusson, Sindri and Bostr{\"o}m, Henrik},
  booktitle={The 13th Symposium on Conformal and Probabilistic Prediction with Applications},
  pages={195--217},
  year={2024},
  organization={PMLR}
}

@incollection{koziel2022fundamentals,
  title={Fundamentals of data-driven surrogate modeling},
  author={Koziel, Slawomir and Pietrenko-Dabrowska, Anna},
  booktitle={Surrogate Modeling for High-Frequency Design: Recent Advances},
  pages={1--37},
  year={2022},
  publisher={World Scientific}
}

@article{lei2018distribution,
  title={Distribution-free predictive inference for regression},
  author={Lei, Jing and G’Sell, Max and Rinaldo, Alessandro and Tibshirani, Ryan J and Wasserman, Larry},
  journal={Journal of the American Statistical Association},
  volume={113},
  number={523},
  pages={1094--1111},
  year={2018},
  publisher={Taylor \& Francis}
}

@article{li2025new,
  title={A new {Bayesian} sparse polynomial chaos expansion based on {Gaussian} scale mixture prior model},
  author={Li, Yangtian and Luo, Yangjun and Zhong, Zheng and Yu, Jing and Wang, Ruiting},
  journal={Mechanical Systems and Signal Processing},
  volume={229},
  pages={112511},
  year={2025},
  publisher={Elsevier}
}

@article{loukrezis2026conformalized,
author = {Loukrezis, Dimitrios and Giovanis, Dimitris G.},
title = {Conformalized polynomial chaos expansion for uncertainty-aware surrogate modeling},
journal = {GAMM-Mitteilungen},
volume = {49},
number = {4},
pages = {e70017},
year = {2026}
}

@article{loukrezis2025multivariate,
  title={Multivariate sensitivity-adaptive polynomial chaos expansion for high-dimensional surrogate modeling and uncertainty quantification},
  author={Loukrezis, Dimitrios and Diehl, Eric and De Gersem, Herbert},
  journal={Applied Mathematical Modelling},
  volume={137},
  pages={115746},
  year={2025},
  publisher={Elsevier}
}

@article{loukrezis2022power,
  title={Power module heat sink design optimization with ensembles of data-driven polynomial chaos surrogate models},
  author={Loukrezis, Dimitrios and De Gersem, Herbert},
  journal={e-Prime-Advances in Electrical Engineering, Electronics and Energy},
  volume={2},
  pages={100059},
  year={2022},
  publisher={Elsevier}
}

@article{loukrezis2020robust,
  title={Robust adaptive least squares polynomial chaos expansions in high-frequency applications},
  author={Loukrezis, Dimitrios and Galetzka, Armin and De Gersem, Herbert},
  journal={International Journal of Numerical Modelling: Electronic Networks, Devices and Fields},
  volume={33},
  number={6},
  pages={e2725},
  year={2020},
  publisher={Wiley Online Library}
}

@article{luethen2022automatic,
  title={Automatic selection of basis-adaptive sparse polynomial chaos expansions for engineering applications},
  author={L{\"u}then, Nora and Marelli, Stefano and Sudret, Bruno},
  journal={International Journal for Uncertainty Quantification},
  volume={12},
  number={3},
  year={2022},
  publisher={Begel House Inc.}
}

@article{manfredi2025hybrid,
  title={A hybrid polynomial chaos expansion--{Gaussian} process regression method for {Bayesian} uncertainty quantification and sensitivity analysis},
  author={Manfredi, Paolo},
  journal={Computer Methods in Applied Mechanics and Engineering},
  volume={436},
  pages={117693},
  year={2025},
  publisher={Elsevier}
}

@article{manfredi2026polynomial,
  title={Polynomial chaos vs kernel machine learning methods for uncertainty quantification: A comparative study and benchmarking with the hybrid {PCE-GPR} approach},
  author={Manfredi, Paolo},
  journal={Computer Methods in Applied Mechanics and Engineering},
  volume={449},
  pages={118523},
  year={2026},
  publisher={Elsevier}
}

@article{marelli2018active,
  title={An active-learning algorithm that combines sparse polynomial chaos expansions and bootstrap for structural reliability analysis},
  author={Marelli, Stefano and Sudret, Bruno},
  journal={Structural Safety},
  volume={75},
  pages={67--74},
  year={2018},
  publisher={Elsevier}
}

@article{matiz2020conformal,
  title={Conformal prediction based active learning by linear regression optimization},
  author={Matiz, Sergio and Barner, Kenneth E},
  journal={Neurocomputing},
  volume={388},
  pages={157--169},
  year={2020},
  publisher={Elsevier}
}

@article{migliorati2014analysis,
  title={Analysis of discrete $L^2$ projection on polynomial spaces with random evaluations},
  author={Migliorati, Giovanni and Nobile, Fabio and Von Schwerin, Erik and Tempone, Ra{\'u}l},
  journal={Foundations of Computational Mathematics},
  volume={14},
  number={3},
  pages={419--456},
  year={2014},
  publisher={Springer}
}

@article{moustapha2022multi,
  title={Multi-objective robust optimization using adaptive surrogate models for problems with mixed continuous-categorical parameters},
  author={Moustapha, Maliki and Galimshina, Alina and Habert, Guillaume and Sudret, Bruno},
  journal={Structural and Multidisciplinary Optimization},
  volume={65},
  number={12},
  pages={357},
  year={2022},
  publisher={Springer}
}

@article{moustapha2019surrogate,
  title={Surrogate-assisted reliability-based design optimization: a survey and a unified modular framework},
  author={Moustapha, Maliki and Sudret, Bruno},
  journal={Structural and Multidisciplinary Optimization},
  volume={60},
  number={5},
  pages={2157--2176},
  year={2019},
  publisher={Springer}
}

@inproceedings{nouretdinov2001ridge,
  title={Ridge regression confidence machine},
  author={Nouretdinov, Ilia and Melluish, Thomas and Vovk, Volodya},
  booktitle={Proceedings of the Eighteenth International Conference on Machine Learning},
  pages={385--392},
  year={2001}
}

@article{oladyshkin2012data,
  title={Data-driven uncertainty quantification using the arbitrary polynomial chaos expansion},
  author={Oladyshkin, Sergey and Nowak, Wolfgang},
  journal={Reliability Engineering \& System Safety},
  volume={106},
  pages={179--190},
  year={2012},
  publisher={Elsevier}
}

@article{owen2017comparison,
  title={Comparison of surrogate-based uncertainty quantification methods for computationally expensive simulators},
  author={Owen, Nicola E and Challenor, Peter and Menon, Prathyush P and Bennani, Samir},
  journal={SIAM/ASA Journal on Uncertainty Quantification},
  volume={5},
  number={1},
  pages={403--435},
  year={2017},
  publisher={SIAM}
}

@inproceedings{papadopoulos2002inductive,
  title={Inductive confidence machines for regression},
  author={Papadopoulos, Harris and Proedrou, Kostas and Vovk, Volodya and Gammerman, Alex},
  booktitle={European Conference on Machine Learning},
  pages={345--356},
  year={2002},
  organization={Springer}
}

@inproceedings{papadopoulos2008normalized,
  title={Normalized nonconformity measures for regression conformal prediction},
  author={Papadopoulos, Harris and Gammerman, Alex and Vovk, Volodya},
  booktitle={Proceedings of the IASTED International Conference on Artificial Intelligence and Applications (AIA 2008)},
  pages={64--69},
  year={2008}
}

@article{papadopoulos2011reliable,
  title={Reliable prediction intervals with regression neural networks},
  author={Papadopoulos, Harris and Haralambous, Haris},
  journal={Neural Networks},
  volume={24},
  number={8},
  pages={842--851},
  year={2011},
  publisher={Elsevier}
}

@book{rao2006probability,
  title={Probability Theory with Applications},
  author={Rao, Malempati Madhusudana and Swift, Randall J},
  year={2006},
  publisher={Springer}
}

@book{rasmussen2005gaussian,
    author = {Rasmussen, Carl Edward and Williams, Christopher K. I.},
    title = {Gaussian Processes for Machine Learning},
    publisher = {The MIT Press},
    year = {2005}
}

@article{romano2019conformalized,
  title={Conformalized quantile regression},
  author={Romano, Yaniv and Patterson, Evan and Candes, Emmanuel},
  journal={Advances in Neural Information Processing Systems},
  volume={32},
  year={2019}
}

@article{rumsey2026bayesian,
  title={Bayesian adaptive polynomial chaos expansions},
  author={Rumsey, Kellin N and Francom, Devin and Gibson, Graham and Derek Tucker, J and Huerta, Gabriel},
  journal={Stat},
  volume={15},
  number={1},
  pages={e70151},
  year={2026},
  publisher={Wiley Online Library}
}

@article{schobi2015polynomial,
  title={Polynomial-chaos-based Kriging},
  author={Schobi, Roland and Sudret, Bruno and Wiart, Joe},
  journal={International Journal for Uncertainty Quantification},
  volume={5},
  number={2},
  year={2015},
  publisher={Begel House Inc.}
}

@book{seber2003linear,
  title={Linear regression analysis},
  author={Seber, George A F and Lee, Alan J},
  year={2003},
  publisher={John Wiley \& Sons}
}

@article{shang2024active,
  title={Active learning of ensemble polynomial chaos expansion method for global sensitivity analysis},
  author={Shang, Xiaobing and Wang, Lipeng and Fang, Hai and Lu, Lingyun and Zhang, Zhi},
  journal={Reliability Engineering \& System Safety},
  volume={249},
  pages={110226},
  year={2024},
  publisher={Elsevier}
}

@article{shao2017bayesian,
  title={Bayesian sparse polynomial chaos expansion for global sensitivity analysis},
  author={Shao, Qian and Younes, Anis and Fahs, Marwan and Mara, Thierry A},
  journal={Computer Methods in Applied Mechanics and Engineering},
  volume={318},
  pages={474--496},
  year={2017},
  publisher={Elsevier}
}

@article{soize2004physical,
  title={Physical systems with random uncertainties: chaos representations with arbitrary probability measure},
  author={Soize, Christian and Ghanem, Roger},
  journal={SIAM Journal on Scientific Computing},
  volume={26},
  number={2},
  pages={395--410},
  year={2004},
  publisher={SIAM}
}

@article{sudret2008global,
  title={Global sensitivity analysis using polynomial chaos expansions},
  author={Sudret, Bruno},
  journal={Reliability Engineering \& System Safety},
  volume={93},
  number={7},
  pages={964--979},
  year={2008},
  publisher={Elsevier}
}

@article{tibshirani2019conformal,
  title={Conformal prediction under covariate shift},
  author={Tibshirani, Ryan J and Foygel Barber, Rina and Candes, Emmanuel and Ramdas, Aaditya},
  journal={Advances in Neural Information Processing Systems},
  volume={32},
  year={2019}
}

@article{torre2019data,
  title={Data-driven polynomial chaos expansion for machine learning regression},
  author={Torre, Emiliano and Marelli, Stefano and Embrechts, Paul and Sudret, Bruno},
  journal={Journal of Computational Physics},
  volume={388},
  pages={601--623},
  year={2019},
  publisher={Elsevier}
}

@book{vovk2005algorithmic,
  title={Algorithmic learning in a random world},
  author={Vovk, Vladimir and Gammerman, Alexander and Shafer, Glenn},
  year={2005},
  publisher={Springer}
}

@inproceedings{vovk2012conditional,
  title={Conditional validity of inductive conformal predictors},
  author={Vovk, Vladimir},
  booktitle={Asian Conference on Machine Learning},
  pages={475--490},
  year={2012},
  organization={PMLR}
}

@article{vovk2015cross,
  title={Cross-conformal predictors},
  author={Vovk, Vladimir},
  journal={Annals of Mathematics and Artificial Intelligence},
  volume={74},
  number={1},
  pages={9--28},
  year={2015},
  publisher={Springer}
}

@article{wang2018fast,
  title={A fast and efficient conformal regressor with regularized extreme learning machine},
  author={Wang, Di and Wang, Ping and Shi, Junzhi},
  journal={Neurocomputing},
  volume={304},
  pages={1--11},
  year={2018},
  publisher={Elsevier}
}

@article{xiu2002wiener,
  title={The {Wiener}--{Askey} polynomial chaos for stochastic differential equations},
  author={Xiu, Dongbin and Karniadakis, George Em},
  journal={SIAM Journal on Scientific Computing},
  volume={24},
  number={2},
  pages={619--644},
  year={2002},
  publisher={SIAM}
}
\end{document}